\documentclass[fleqn,usenatbib]{mnras}

\usepackage{newtxtext,newtxmath}
\usepackage[T1]{fontenc}
\usepackage{floatpag}
\usepackage{adjustbox}
\usepackage{subcaption}
\usepackage{graphicx}
\usepackage{array}
\usepackage{amsmath}
\usepackage{nicefrac}
\usepackage{bm}
\usepackage{tabularx}
\floatpagestyle{empty}

\DeclareRobustCommand{\VAN}[3]{#2}
\let\VANthebibliography\thebibliography
\def\thebibliography{\DeclareRobustCommand{\VAN}[3]{##3}\VANthebibliography}

\usepackage{longtable}
\usepackage{graphicx}	% Including figure files
\usepackage{amsmath}	% Advanced maths commands
\usepackage{subcaption}
\usepackage[version=3]{mhchem} %Package for chemcial symbols
\usepackage[linesnumbered,ruled,vlined]{algorithm2e}
\title[PTDEs:The elixir of life]{Partial tidal disruption events: The elixir of life}

\author[M. Sharma et al.]{
Megha Sharma,\thanks{E-mail: megha.sharma@monash.edu}
Daniel J. Price,
and Alexander Heger
\\
School of Physics \& Astronomy,  Monash University, Vic. 3800, Australia\\
}

\date{Accepted XXX. Received YYY; in original form ZZZ}

\pubyear{2024}

\begin{document}
\label{firstpage}
\pagerange{\pageref{firstpage}--\pageref{lastpage}}
\maketitle

% Abstract of the paper
\begin{abstract}In our Galactic Center, about $10\mathord,000$ to $100\mathord,000$ stars are estimated to have survived tidal disruption events, resulting in partially disrupted remnants.  These events occur when a supermassive black hole (SMBH) tidally interacts with a star, but not enough to completely disrupt the star.  We use the 1D stellar evolution code \textsc{Kepler} and the 3D smoothed particle hydrodynamics code \textsc{Phantom} to model the tidal disruption of $1\,\mathrm{M}_\odot$, $3\,\mathrm{M}_\odot$, and $10\,\mathrm{M}_\odot$ stars at zero-age (ZAMS), middle-age (MAMS), and terminal-age main-sequence (TAMS).  We map the disruption remnants into \textsc{Kepler} in order to understand their post-distribution evolution. We find distinct characteristics in the remnants, including increased radius, rapid core rotation, and differential rotation in the envelope.  The remnants undergo composition mixing that affects their stellar evolution.  Whereas the remnants formed by disruption of ZAMS models evolve similarly to unperturbed models of the same mass, for MAMS and TAMS stars, the remnants have higher luminosity and effective temperature. Potential observational signatures include peculiarities in nitrogen and carbon abundances, higher luminosity, rapid rotation, faster evolution, and unique tracks in the Hertzsprung-Russell diagram. 
\end{abstract}

% Select between one and six entries from the list of approved keywords.
% Don't make up new ones.

%%%%%%%%%%%%%%%%%%%%%%%%%%%%%%%%%%%%%%%%%%%%%%%%%%

%%%%%%%%%%%%%%%%% BODY OF PAPER %%%%%%%%%%%%%%%%%%
%https://ui.adsabs.harvard.edu/search/q=citations(bibcode%3A2009MNRAS.397.2148G)&sort=citation_count%20desc%2C%20bibcode%20desc&p_=0
\section{Introduction}

\citet{TalandMario2001} argued that the Galactic Centre of the Milky Way harbours about $10^4$--$10^5$ stars that survived a tidal disruption event (TDE) with the supermasive black hole (SMBH) at its centre.  \citet{Manukian2013} estimated a similar number of stars. The upcoming Vera C. Rubin Legacy Survey of Space and Time will result in an explosion in the detection of partial TDEs \citep{Hambleton2023}.  Understanding partial TDEs hence is essential.

Supermassive black holes exist at the centre of most galaxies \citep{Miyoshi1995,Macchetto1997,Ghez1998,Ghez2008,EHTC2019}.  Our Galaxy hosts a $\sim4\times10^6\,\mathrm{M}_\odot$ SMBH (Sgr A$^*$; \citealt{Ghez1998,Do2019,Gravity2020,EHTC2022}).  SMBHs can be surrounded by nuclear star clusters (NSCs), dense conglomerates of stars \citep{Kormendy1988,Genzel1994,Genzel2010,Georgiev2014}.  Within the mass range of $10^9\,\mathrm{M}_\odot$ to $10^{10}\,\mathrm{M}_\odot$, the NSC and SMBH coexist \citep{Filippenko2003,Seth2008,Neumayer2020}.  This coexistence may be indicative of a co-evolutionary process between the SMBH and the NSC \citep{Graham2009,Kormendy2013,Antonini2015}.  For galaxies with $M > 10^{11}\,\mathrm{M}_\odot$, there is weak evidence of the presence of NSCs alongside SMBHs.  Evidence supporting the existence of central massive black holes in galaxies with masses below $10^9\,\mathrm{M}_\odot$ is also limited, although a few candidates have been identified \citep{Neumayer2020}.

NSCs are home to stars with a range of ages and metallicity \citep{Neumayer2020}. 
 Our closest SMBH, Sgr A* co-exists with a NSC \citep{Becklin1968,Balick1974} of mass $\sim 2$--$4 \times 10^7\,\mathrm{M}_\odot$ (see references within \citealt{Neumayer2020}). 
%The proximity of this NSC compared with other galaxies makes it the best laboratory for understanding the population of NSCs \citep{Schodel2014}.

Most of the stars in our NSC are old (age $>5\,\mathrm{Gyr}$), late-type giant stars and helium-burning stars on the horizontal branch with mass between $0.5\,\mathrm{M}_\odot$ to $4\,\mathrm{M}_\odot$.  A few asymptotic giant branch stars with temperatures less than $2\mathord,700\,\mathrm{K}$ along with a few red super-giant stars have also been observed.  About $200$ young ($3$--$8\,\mathrm{Myr}$) Wolf-Rayet, O and B type stars have been detected within $0.5\,\mathrm{pc}$.  Possible explanations of their existence include short events or weak starbursts about $6$ Myr ago or the migration of a star cluster to the central parsec (see reviews by \citealt{Genzel2010,Neumayer2020}).  Within $0.04\,\mathrm{pc}$, a cluster of B-stars orbiting Sgr A* on highly eccentric and inclined orbits have been detected, called the S-star cluster.  Their existence in such proximity to the SMBH is puzzling, with proposed explanations including the Hills mechanism \citep{Hills1988} and migration \citep{Genzel2010}.  About six G-objects --- having characteristics associated with dust and gas clouds, but dynamics similar to stellar-mass objects \citep{Ciurlo2020}, have also been detected within $0.04\,\mathrm{pc}$  \citep{Gillessen2012,Phifer2013,Witzel2017,Ciurlo2020}.  The cause of their formation remains a mystery.

The dynamics of the stars present in NSC are influenced by the SMBH's radius of influence ($r_{\rm h}=GM_\bullet/\sigma^2$), where $M_\bullet$ is the mass of SMBH and $\sigma$ is the velocity dispersion.  Gravitational relaxation processes --- perturbations by massive objects \citep{Perets2007}, and/or two-body stellar relaxation interactions \citep{Magorrian1999} result in stars entering the loss cone.  The loss cone is defined as a region in the phase space where the orbits have the closest approach to the black hole (pericentre; $r_{\rm p}$) less than the tidal radius ($r_{\rm t}$) given by 
\begin{equation}
\label{eq:tidal radius}
        r_\mathrm{t} = 0.47 \;\mathrm{au}\; \left(\frac{M_{\mathrm{\bullet}}}{10^6\; \mathrm{M_\odot}}\right)^{1/3}\left(\frac{M_{*}}{\mathrm{M_\odot}}\right)^{-1/3}\;\left(\frac{R_{*}}{\mathrm{R_\odot}}\right)\;,
\end{equation}
where $M_{\mathrm{*}}$ is the stellar mass and $R_{*}$ is the radius of the star \citep{Magorrian1999}.  The fate of these stars can include tidal disruption by the SMBH, direct plunge into SMBH, capture by SMBH and inspiral, or ejection \citep{Hills1975,Tal2017}.  Not all stars that enter the loss cone are disrupted by the SMBH, as they can move in and out of the loss cone several times per orbit \citep{Magorrian1999}. 

Direct $N$-body simulations suggest that almost all stars entering the tidal radius have eccentricities close to unity, with about half of the stars having $e > 1$, implying marginally hyperbolic orbits \citep{Zhong2014}.  \citet{Tal2017} argued that the stars scattered into the loss cone are on hyperbolic orbits with the specific orbital energy a small fraction of the binding energy. Hence, these orbits can be approximated as parabolic orbits.  \citet{Hayasaki2018} found, using $N$-body simulations, that stars that are disrupted by the SMBH are rarely on an eccentric, precisely parabolic or hyperbolic orbit.  Most of the particles followed marginally eccentric or marginally hyperbolic orbits.  \citet{Zhong2023} also reported a similar result.

We focus on partial tidal disruption events (TDEs) in this paper.  A star is tidally disrupted when it approaches the tidal radius, a distance at which the gravitational forces of the SMBH can overcome the self-interaction forces of a star \citep{Hills1975,Rees1988,Evans1989}.  TDEs were theorised in the 1970s \citep{Hills1975} and first observationally detected in the 1990s by the ROSAT All Sky Survey in X-rays \citep{Donely2002}.

During a TDE, a star can either undergo a full disruption or leave behind a remnant --- a partial TDE.  The strength of a TDE is quantified by the penetration factor $\beta \equiv r_\mathrm{t}/r_\mathrm{p}$ \citep{Lodato2009,Guillochon2013}.  The boundary between full and partial disruption remains uncertain due to several factors such as the star's structure, SMBH spin and general relativistic effects \citep{Gezari2021}.

A few dozen TDEs have been detected in UV \citep{Martin2005}, optical \citep{VanV2011}, soft and hard X-rays since the first detection (see \citealt{Gezari2021} and references within).  A TDE rate of $10^{-4}$--$10^{-5}$ per galaxy per year \citep{Magorrian1999,Brockamp2011,Stone2016} has been estimated but this can be influenced by the spin of the SMBH, the nature of the galaxy (e.g. post-starburst galaxies have a higher rate \citealt{StoneVan2016}) and the presence of binary SMBH (see review by \citep{Gezari2021}).  In the past few years, a few partial TDEs have been inferred based on the decay rate of their lightcurves \citep{Blagorodnova2017,Nicholl2020}.  Recently, repeating partial TDEs have also been observed \citep{Payne2021,Miniutti2023,Huang2023,Somalwar2023,Wevers2023}. 

\citet{Stone2020} argued that partial TDEs should be more common than full disruptions due to the larger cross-section.  \citet{Chen2021} found that partial TDEs become dominant for $M_\bullet \geq 10^6\,\mathrm{M}_\odot$.  Using $N$-body simulations, \citet{Zhong2022} suggested that partial TDEs are $75\%$ more likely than analytical estimates based on the encounter cross-section, as a single star can undergo repeated partial TDEs.  Recently, \citet{Bortolas2023} found that partial TDEs are more likely than full TDEs by a factor of $\sim 10$. 

Previous studies have delved into understanding the properties of remnants formed during partial TDEs.  \citet{TalandMario2001} analytically explored remnants of a $\gamma = 1.5$ polytrope and a solar-type star.  Both models were initially on slightly hyperbolic orbits in Newtonian gravity. They argued that stars would experience spin-up from the SMBH resulting in rotationally induced mixing.  The remnants would be bluer and more luminous due to the stripping of material compared with a main-sequence star of the same mass.  Hydrodynamical simulations by \citet{Antonini2011} indicate that stars bound to SMBH can undergo multiple TDEs and result in depletion of the outermost stellar region, leaving behind a hot central core. 

\citet{Manukian2013} used simulations of partial disruptions of solar-type stars on parabolic orbits, finding that stars can have velocities of several hundred $\mathrm{km}\,\mathrm{s}^{-1}$, preventing them from escaping the galactic centre. \citet{Goicovic2019} simulated the tidal disruption of a $1\,\mathrm{M}_\odot$ zero-age main-sequence (ZAMS) star and found that the surviving core has positive angular momentum in outer layers, while negative in the innermost region. 

\citet{Guillochon2013} explored the boundary between partial and full TDEs for polytropic stellar models of $\gamma = \nicefrac43$ and $\gamma = \nicefrac53$ using an adaptive-mesh grid-based hydrodynamics code.  They found a critical $\beta$ of $1.85$ for $\gamma = \nicefrac43$ and $0.9$ for $\gamma = \nicefrac53$. 
 \citet{Goicovic2019} determined a critical $\beta \sim 2.5$ for their $1\,\mathrm{M}_\odot$ ZAMS model.  \citet{Lawsmith2020} performed a detailed study with a range of stellar masses for stellar models at different ages.  They found that critical $\beta$ is related to the mean ratio of the central and mean density of a star ($\rho_\mathrm{c}/\bar{\rho}$).  \citet{Ryu32020} utilised a fully general relativistic framework for a range of stellar masses at middle age main-sequence (MAMS), discovering remnants with velocities of few hundred km/s for strong disruptions (mass of remnant $\approx 40$ to $60\%$ of the initial star mass). Bound remnants were all on highly eccentric orbits.  They found remnants of angular frequencies close to break-up in the outer layers. 

In this paper, we study how the stellar evolution of disrupted remnants proceeds post disruption.  To address this question, 
we evolve stars in a 1D stellar evolution code, \textsc{Kepler} \citep{Weaver1978}, disrupt them using a 3D smoothed particle hydrodynamics code, \textsc{Phantom} \citep{Price2018}, considering general relativistic effects \citep{Liptai2019} with Kerr metric \citep{Kerr1963} of zero spin, and map them back into 1D to continue their lives.  We explore the disruption of stars with masses of $1\,\mathrm{M}_\odot$, $3\,\mathrm{M}_\odot$, and $10\,\mathrm{M}_\odot$ at three different stages of their lives for a range of penetration factors.  A key finding is that disrupted stars are `forever young', staying on the main-sequence for longer than the age of the Universe. For stars, partial disruptions are the long-sought Holy Grail, the fountain of youth and the elixir of life.

This paper is structured as follows: In Section~\ref{sec:Method_sec} we describe the method used to simulate TDEs and map the resulting remnants back into \textsc{Kepler} to continue their evolution.  In Section~\ref{sec:results_sec} we present our results.  In Section~\ref{sec:discussion_sec} we discuss our results and compare them with current literature.  We summarise in Section~\ref{sec:conclusion_sec}.  

\section{Methods}
\label{sec:Method_sec}
We used the general relativistic smoothed particle hydrodynamics code \textsc{Phantom} \citep{Price2018,Liptai2019}, and the stellar evolution code \textsc{Kepler} \citep{Weaver1978} for our simulations. We mapped \textsc{Kepler} models into \textsc{Phantom} to simulate the tidal disruption, and then remapped into \textsc{Kepler} to continue the evolution. 

\subsection{Mapping from 1D to 3D}
\begin{figure}
	\includegraphics[width=\columnwidth]{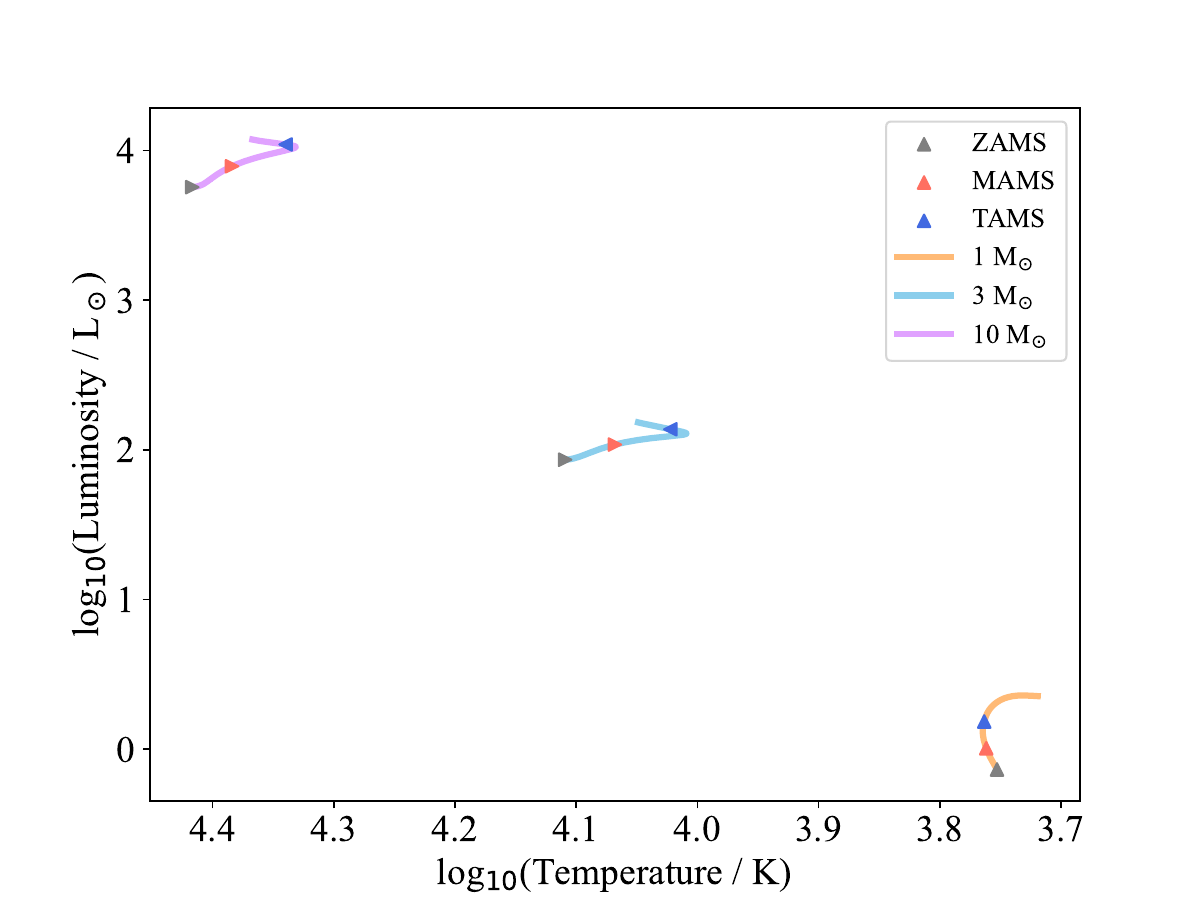}
    \caption{Hertzsprung-Russell (H-R) diagram of the $1\,\mathrm{M}_\odot$, $3\,\mathrm{M}_\odot$, and $10\,\mathrm{M}_\odot$ stars we evolved in \textsc{Kepler}.  We also show the three different stages (ZAMS, MAMS, and TAMS) of the stars for which the models were obtained. }
    \label{fig:Hr diagram}
\end{figure}

We computed stellar models using \textsc{Kepler} for stars with masses of $1\,\mathrm{M}_\odot$, $3\,\mathrm{M}_\odot$, and $10\,\mathrm{M}_\odot$ at three different stages of the star's lifecycle.  
 Table~\ref{tab:models_used} lists the parameters of our models. These stages correspond to the Zero-Age Main-Sequence (ZAMS), Middle-Age Main-Sequence (MAMS), and Terminal-Age Main-Sequence (TAMS) of the star.  ZAMS, MAMS, and TAMS are defined as the stellar evolution stages when the mass fraction of hydrogen in the core has dropped by $1\%$ (absolute), is half the initial value, and when it has reached $1\%$ (absolute), respectively.  Figure~\ref{fig:Hr diagram} shows the evolutionary paths of these \textsc{Kepler} models.  We also show the three stages at which the models were analysed.  As the star evolves on the main-sequence, the luminosity increases while the effective temperature decreases.
Our \textsc{Kepler} models use an approximate nuclear network, \texttt{APPROX19}, that comprises neutrons, \ce{^1H}, \ce{^3He}, \ce{^4He}, \ce{^12C}, \ce{^14N}, \ce{^16O}, \ce{^20Ne}, \ce{^24Mg}, \ce{^28Si}, \ce{^32S}, \ce{^36Ar}, \ce{^40Ca}, \ce{^44Ti}, \ce{^48Cr}, \ce{^52Fe}, \ce{^54Fe}, and \ce{^56Ni}.  This network is adequate to capture the main nuclear burning phases of massive stars and their nuclear energy generation. 

To obtain a 3D density profile from the initial 1D density data, we used the stellar profile mapping and relaxation procedure outlined in \citet{Lau2022}.  We fixed the radial entropy profile of the star during the relaxation.  We stopped the relaxation procedure when the ratio of kinetic to potential energy dropped below $10^{-7}$ and the error in density was less than $1\%$.  All stellar models were resolved with $10^6$ SPH particles.  We employed a specific heat ratio of $\gamma=\nicefrac{5}{3}$ and used an adiabatic equation of state to calculate the energy and temperature from the relaxed pressure and density profiles. 

The composition data obtained from \textsc{Kepler} was then interpolated onto the SPH particle positions and saved into a file.  We did not simulate nuclear reactions in \textsc{Phantom}, as the simulation duration was significantly shorter than the nuclear-burning timescales even during the peak of the compression.

\begin{table}
\centering
 \caption{Properties of the nine \textsc{Kepler} models used in this paper. Columns 1 and 2 give initial mass and evolutionary stages, respectively. ZAMS, MAMS, and TAMS are zero-age, middle-age, and terminal-age main-sequence. Column 3 gives the age at which the tidal disruption occurs. Columns 4 and 5 list the central density and radius of the model respectively.  Column 6 gives the ratio of the central density to the mean density.  The last column lists the critical $\beta$, where no remnant would be formed calculated by Section~\ref{sec:critical_beta_sec}.  The uncertainties are based on the coarseness of the simulations run. }
 \label{tab:models_used}
 \begin{tabular}{ccrrrrr}
  \hline
    \hline 
  Mass & Stage & Age & $\rho_\mathrm{c}$ & R  & $\rho_\mathrm{c}$/$\bar{\rho}$ & $\beta_\mathrm{c}$ \\
  ($\mathrm{M}_\odot$) & &  ($\mathrm{Myr}$) & ($\mathrm{g/cm}^3$) &  ($\mathrm{R}_\odot$) &  &  \\
  \hline

  1 & ZAMS & 150 & 82.75 & 0.89 & 41.49 & 1.56$\pm$0.1\\
  1 & MAMS & 4,530 & 151.02 & 1.01 & 110.67 & 2.04$\pm$0.1\\
  1 & TAMS & 8,620 & 430.47 & 1.23 & 569.76 & 3.40$\pm$0.5\\
  \hline
  3 & ZAMS & 9.2 & 44.45 & 1.87 & 68.91 & 1.91$\pm$0.1\\
  3 & MAMS & 175.9 & 44.03 & 2.54 & 171.06 & 2.59$\pm$0.2\\
  3 & TAMS & 283.8 & 87.12 & 3.53 & 908.56 & 4.80$\pm$0.3\\
   \hline
    10 & ZAMS & 0.38 & 10.41 & 3.69 & 37.20 & 1.59$\pm$0.1\\
  10 & MAMS & 12.4 & 9.85 & 5.04 & 89.33 & 2.17$\pm$0.3\\
  10 & TAMS & 19.9 & 20.29 & 7.31 & 563.72 & 4.49$\pm$0.3\\
  \hline
   \hline 
 \end{tabular}
\end{table}
% First figure is the HR diagram and show the 3 models used 

\subsection{Disrupting stars in \textsc{Phantom}}
Once the star was relaxed, we placed it onto an orbit corresponding to a Newtonian parabolic orbit around a $10^6\, \mathrm{M}_\odot$ SMBH at a distance $\sim 10$ times the tidal radius (Eq.~\ref{eq:tidal radius}).  We later realised that this procedure resulted in slightly hyperbolic orbits due to the difference between the general relativistic and Newtonian energy at $10\,\mathrm{r}_\mathrm{t}$.  While this difference results in a faster velocity of the remnant at infinity, the effect on the tidal disruption of the star is not significant (see Appendix~\ref{app:zero_e_orbits} for a comparison).

We used the following conditions in our simulations in \textsc{Phantom}.  A Courant number of $0.3$ was used which is also the default value. As we used equal mass particles, we used a proportionality factor of $1.2$ that specified the smoothing length in terms of mean local particle spacing.  To use the smoothing length and density interchangeably during the time-stepping, we used a tolerance of $10^{-4}$ which is the default value.  A tolerance of $10^{-7}$ was used for both position and momentum iterations in the GR timestep integrator \citep{Liptai2019}.  We also used a tree accuracy of $0.5$ in the self-gravity calculation. We used high-resolution shock-capturing dissipation in our simulations, with the shock viscosity factors $\alpha=1$ and $\beta=2$.  We also used a shock conductivity of $0.1$. We used an adiabatic equation of state with a mean molecular weight of $0.5988$ for all the models.  Heating from $P\,\mathrm{d}V$ work and shocks were assumed to be trapped rather than radiated. We used $K=P/\rho^\gamma$ as the energy variable \citep{Liptai2019}.  We performed simulations in the \citet{Kerr1963} metric with an SMBH with no spin \citep{Liptai2019}. 

We performed the initial disruption with an accretion radius set to the last stable orbit at $6\,GM/c^2$ inside of which particles were deleted from the simulation. 
 After the initial disruption, for computational efficiency, we manually reset the accretion radius such that the material falling onto the SMBH would be removed but not so large that it would remove the material falling back onto the remnant (about $100$--$1\mathord,000\,GM/c^2$ depending on the simulation).

 We assumed that the mass of the accretion disk formed after disruption is negligible compared to the mass of the SMBH and the particles within the accretion radius are accreted by the SMBH.  

Table~\ref{tab:all_sims_data} lists the $\beta$ values, pericentre distance of our simulations.

\begin{table}

\centering
\caption{Properties of \textsc{Phantom} simulations.  Column 1 lists the model evolution stage; Column 2, the impact parameter, $\beta$; Column 3, the mass of the remnant; Column 4 and 5, the star's initial and the remnant's final (error $\pm 100$ km/s) escape velocities; Column 6, the Newtonian pericentre,  $r_\mathrm{p}=r_\mathrm{t}/\beta$; and Column 7, the Kelvin-Helmholtz time of the remnant.   }
 %\begin{tabular}{c|c|c|c|c|c}
% \begin{adjustbox}{width=\textwidth}
\small
\label{tab:models_run}
\begin{tabular}{crrrrrr}
% \begin{tabularx}{\textwidth}{|X|X|X|X|X|X|X|X|}

  \hline
    \hline
%  Stage & $\beta$ & M$_{\rm{rem}}$ &  v$_\infty$ & v$_\infty$ & r$_{\mathrm{p}}$ & t$_\mathrm{KH}$ of  \\
%     &  &   & initial & final &  &  remnant  \\
  Stage & $\beta$ & $M_{\rm{rem}}$ &  $v_{\infty,\mathrm{ini}}$ & $v_{\infty,\mathrm{fin}}$ & $r_{\mathrm{p}}$ & $t_\mathrm{KH,rem}$ \\
   &  &  ($\mathrm{M}_\odot$) & ($\mathrm{km/s}$)  & ($\mathrm{km/s}$) &($\mathrm{R}_\odot$) & ($\mathrm{Myr}$) \\
%  \hline 
%  \multicolumn{7}{c}{$1\,\mathrm{M}_\odot$}\\
  \hline 
%  \noalign{\smallskip}
$1\,\mathrm{M}_\odot$ 
& 0.53 & 1.00  & 2,034 & 2,030 & 167.43 & 15.6 \\
ZAMS 
& 0.85 & 0.96 & 2,078 & 2,067 &  104.64 & 16.9\\
& 1.06 & 0.84 &  20,93 & 2,081 & 83.72 & 21.6\\
& 1.28 & 0.58 & 2,102 & 2,117 &  69.76 & 42.1\\
& 1.49 & 0.14 & 2,109 & 2,197 &  59.79 & 510.2\\
  \hline
$1\,\mathrm{M}_\odot$ 
& 0.54 & 1.00 &  1,813 & 1,809 & 187.35 & 15.6\\
 MAMS
& 0.86 & 0.97 &  1,853 & 1,848 & 117.10 & 16.4 \\
& 1.08 & 0.91 & 1,865 & 1,817 &  93.68 & 18.6\\
& 1.29 & 0.70 & 1,874 & 1,857 &  78.06 & 30.0\\
& 1.51 & 0.59 & 1,879 & 1,889 &  66.91 & 39.9\\
& 1.72 & 0.37 & 1,884 & 1,893 &  58.55 & 93.5\\
& 1.94 & 0.14 & 1,888 & 1,907 &  52.04 & 557.7\\
   \hline
$1\,\mathrm{M}_\odot$ 
& 0.54 & 1.00 &  1,482 & 1,481 & 227.95 & 15.6\\
 TAMS
& 1.08 & 0.94 & 1,524 & 1,482 &  113.98 & 17.3\\
& 1.62 & 0.75 & 1,538 & 1,502 &  75.98 & 25.9\\
& 2.16 & 0.47& 1,545 & 1,523 &  56.99 & 59.9\\
& 2.70 & 0.24 & 1,549 & 1,478 &  45.59 & 203.8\\
& 3.45 & 0.01 &  1,554 & 1,472 & 35.62 & 3200.6\\
  \hline
%    \multicolumn{7}{c}{$3\,\mathrm{M}_\odot$}\\
%  \hline 
$3\,\mathrm{M}_\odot$ 
& 0.49 & 3.00 & 1,302 & 1,286 & 262.94 & 0.89 \\
  ZAMS
& 0.99 & 2.95 &  1,338 & 1,333 & 131.47 & 0.93\\
& 1.28 & 2.66 & 1,346 & 1,294 &  101.13  & 1.23 \\
& 1.48 & 2.13 & 1,350 & 1,357 &  87.65 & 2.19\\
& 1.78 & 0.64 & 1,354 & 1,576 &  73.04 & 35.17\\
  \hline
$3\,\mathrm{M}_\odot$
& 0.50 & 3.00 & 968 & 945 &  353.98 & 0.89\\
 MAMS 
& 0.80 & 3.00 & 986 & 964 &  221.24 & 0.89\\
& 0.99 & 2.97 &  990 & 989 & 176.99 & 0.92\\
& 1.29 & 2.82 & 998 & 932 &  136.15 & 1.02\\
& 1.49 & 2.61 &  1,000 & 930 & 117.99 & 1.29\\
& 1.79 & 2.06 & 1,001 & 981&  98.33 & 2.39\\
& 2.09 & 1.40 & 1,007 & 999 &  84.28 & 6.52\\
& 2.39 & 0.64 &  1,008 & 1,104 & 73.75 & 34.54\\
   \hline
$3\,\mathrm{M}_\odot$
& 0.50 & 3.00 &  698 & 668 & 489.07 & 0.89 \\
 TAMS 
& 1.00 & 2.98 & 715 & 689 &  244.53 & 0.92\\
& 1.50 & 2.78 &  723 & 608 & 163.02 & 1.09 \\
& 2.00 & 2.31 &  726 & 663 & 122.27 & 1.78 \\
& 2.50 & 1.71 & 728 & 622 &  97.81 & 3.89\\
& 3.00 & 1.26 &  730 & 535 & 81.51 & 8.49 \\
& 4.00 & 0.67 & 729 & 349  &  61.13 & 32.32 \\
& 4.50 & 0.34 & 734 & 384 &  54.34 & 107.11 \\
  \hline 
%    \multicolumn{7}{c}{$10\,\mathrm{M}_\odot$}\\
%  \hline 
$10\,\mathrm{M}_\odot$
& 0.49 & 10.00 & 984 & 970 &  347.41 & 0.04 \\
 ZAMS 
& 0.99 & 9.52 & 1,009 & 964 &  173.70  & 0.04\\
& 1.18 & 7.87 &  1,013 & 1,037 & 144.75 &  0.07\\
& 1.38 & 4.63 & 1,016 & 1,226 & 124.07 &  0.29\\
& 1.48 & 2.53 & 1,019 & 1,380 & 115.80 & 1.39\\
  \hline 
$10\,\mathrm{M}_\odot$
& 0.59 & 9.98 & 718 & 688 &  400.11 & 0.04\\
   MAMS 
& 0.98 & 9.79 & 730 & 701 & 240.06& 0.04 \\
& 1.46 & 7.30 &  737 & 718 & 160.04& 0.09\\
& 1.76 & 4.84 & 737 & 776 &  133.37& 0.3\\
& 1.95 & 2.85 &  738 & 923 & 120.03& 1.02 \\
& 2.15 & 0.27 & 742 & 1,244 &  109.12& 202.49\\
    \hline
$10\,\mathrm{M}_\odot$
& 0.47 & 9.95 & 467 &  414 & 722.78 & 0.04\\
   TAMS 
& 0.94 & 9.88 &  485 & 440 & 361.39 & 0.04 \\
& 1.41 & 9.03 & 486 & 347 &  240.93 & 0.05\\
& 1.88 & 7.16 & 491 & 420 &  180.69 & 0.09\\
& 2.35 & 5.52 & 492 & 209 &  144.56 & 0.18\\
& 2.82 & 4.34 & 492 & (bound) &  120.46 & 0.34\\
& 3.29 & 3.32 & 491 & (bound) &  103.25 & 0.69\\
& 3.76 & 2.21 & 488 & (bound) &  90.35 & 1.98\\
& 4.23 & 0.93 & 496 & 252 &  80.31 & 17.94\\
  \hline 
    \hline  
    \label{tab:all_sims_data}
 \end{tabular}
 % \end{tabularx}
 % \end{adjustbox}  
\end{table}

\begin{table}
\centering
\caption{Semi-major axis and eccentricity of the bound models presented in Table~\ref{tab:all_sims_data}. All bound remnants were formed by disrupting the $10\,\mathrm{M}_\odot$ TAMS model. Column 1 gives the impact parameter $\beta$, Column 2, the semi-major axis, and Column 3, the eccentricity, $e$.  These  Keplerian orbit elements assume Newtonian dynamics only.  
}
\begin{tabular}{ccc}
\hline
 \hline 
$\beta$ & $a$ & $\mathrm{log}_{10}(1-e)$ \\

 & $(\mathrm{AU})$ &  \\
 \hline
 \noalign{\smallskip}
 2.82 & $7.8\times10^3$ &-4.2  \\
 3.29 & $3.8\times10^3$ &-3.8 \\
 3.76 & $4.6\times10^3$ & -3.9 \\
\hline

\end{tabular}

\label{tab:ecc_a_main}
\end{table}

\subsection{Mapping back: 3D to 1D}
To model the post-disruption stellar evolution, we mapped the \textsc{Phantom} models back into \textsc{Kepler}.  To project the 3D data from \textsc{Phantom} back into 1D \textsc{Kepler} models, we binned the star by radius. 

\subsubsection{What is part of a remnant?}
\label{sec:method_rem}
We used the point of maximum density as the centre of the star instead of using the centre of mass of all the particles because the removed particles/debris could move to very large distances, which could result in an inaccurate centre. To determine which particles were part of the debris, we sorted particles by radius with respect to the centre and then calculated the sum of kinetic, potential and internal energy for each particle $i$,
\begin{equation}
    \label{eq:energy_calc}
    E_i = m_i 
    \left(\frac{1}{2}  v_i^2 + \Phi_i +  u_i\right)\;,
\end{equation}
where $\Phi_i$ is the self-gravitional potential between particles calculated by \textsc{Phantom}, $v_i$ is the velocity of the particle relative to the maximum density particle and $m_i$ is the particle mass, which is fixed in \textsc{Phantom}.  We ignored the black hole as the star was sufficiently far away at the time of analysis.  We determined that a particle was bound to the remnant if the total energy was negative and the kinetic energy was less than half of the gravitational potential energy, according to the virial theorem.  This ensured that only particles close to the point of central density were considered as part of the remnant. 

Figure~\ref{fig:temperature_cut} shows that the remnant stars from stronger disruptions contain streams.  By examining the debris temperature as a function of radius, we found that more particles were dispersed horizontally at lower temperatures.  We leveraged this to distinguish particles that were part of the stream from those that were part of the hydrostatic star.  See details in Appendix~\ref{app:sectionA}.

\subsubsection{Binning algorithm}
\label{sec:binning_algo}
We calculated the density, temperature, composition, angular velocity, and radial velocity of each SPH particle.  To determine temperature, we first computed the mean molecular weight based on each particle's composition and then used it along with specific energy in the adiabatic equation of state to find the particle's temperature.

The radial velocity for each particle was determined using
\begin{equation}
    \mathbf{v}_r = \frac{\mathbf{v} \cdot \mathbf{r}}{|\mathbf{r}|}\;,
\end{equation}
where $\mathbf{r}$ was taken with respect to the position of the particle with the maximum density in the remnant. The angular momentum for each particle was calculated using
\begin{equation}
\mathbf{L} = (\mathbf{r} \times \mathbf{v}) \, \mathrm{m}\;.
\end{equation}

We sorted particles by radius and then fixed the number of bins to a constant, $N_{\text{bins}}$.  To achieve optimal mass resolution in the core while limiting the ratio of masses in neighbouring zones to not exceed two, we binned the particles by the following procedure.  We initially placed one particle in the first bin.  Moving outward in radius, we then doubled the number of SPH particles in each bin until the desired mass resolution of $N_\mathrm{p}/N_{\text{bins}} \cdot m_i$, where $N_p$ is the number of SPH particles in the remnant star, and $m_i$ is the particle mass.  We continued adding particles until reaching the desired maximum zone mass of $N_\mathrm{p}/N_{\text{bins}} \cdot m_i$ or the radius is increased by $30\%$, whichever was achieved first. 

Finally, we calculated the required quantities for each bin.  Density, temperature, and radial velocity were determined by averaging all particles within each bin. We set a minimum temperature of $1\mathord,000\,\mathrm{K}$.  
 To compute angular velocity, we computed the moment of inertia tensor for each bin, accounting for variations in axes of rotation among the particles, 
\begin{equation}
    \mathrm{I_{ij}} = \sum_{k=1}^{N} \mathrm{m}_k \; (\ |\mathbf{r}|^2 \; \delta_{ij} \;-r_i^kr_j^k)\;.
\end{equation}
where $I_{ij}$ were the components of the tensor $\mathbf{I}$.
We then obtained the angular velocity. 
\begin{equation}
    \bm{\omega} = \mathbf{I}^{-1}\;\mathbf{L}\;.
\end{equation}
 
\subsubsection{\textsc{Kepler} simulations}
\textsc{Kepler} is a 1D hydrodynamic stellar evolution code and solves the equation of stellar structure by modelling the conservation of energy, mass, and momentum \citep{Weaver1978}.  
% The model implemens dynamical time-stepping and calculates if a shell is thermodynamically stable against convection or not at each iteration. If a shell or zone is stable, there is no convection, and the energy transport is dominated by radiation and other mechanisms. 
Convection, which is inherently multi-dimensional, is modelled using the Mixing Length Theory (MLT; \citealt{Bohm1958}) with a mixing length  of  
\begin{equation}
    l = \alpha H_\mathrm{P} = \alpha \frac{P}{\rho g}\;,
\end{equation}
where $\alpha$ is the mixing length parameter, $P$ is the pressure, $\rho$ is the density, and $g$ is the local gravitational acceleration \citep{Bohm1958}.  The value of $\alpha$ depends on the properties of the convection zone and is well determined for the surface convection zone of our sun, where we can accurately measure effective temperature, luminosity, and even structure from helioseismology.  Both mixing length parameters and opacity scaling were tweaked to give the solar luminosity at the present age of the sun.
%For our \textsc{Kepler} models, given the other default input physics in place, we adopt $\alpha=0.82$ for a $1\,\mathrm{M}_\odot$ model to match the properties of the present-day sun.
As convection does not play a significant role in stars which are $\geq1.3\,\mathrm{M}_\odot$, a value of $\alpha=1$ was used for $3$ and $10\, \mathrm{M}_\odot$ models. 
%We limited the convection velocity to the local sound speed and did not perform convection if the absolute value of the zone velocity exceeded 0.1 times the local sound speed. We also set the convective velocity same as the local sound speed. MLT uses Schwarzschild criteria only and hence, to make realistic models, semiconvection mixing and overshoot were implemented \citep{Renzini1987,Alongi1991,Alogni1993,Freytag1996}.
\textsc{Kepler} uses the Ledoux criterion for convective stability, as well as semiconvection and thermohaline convection \citep{HWS2005}

When mapping models back into \textsc{Kepler} after the binning procedure implemented in \textsc{Phantom}, we first let the model relax until the radius of the star was less than the initial model radius, and the timestep was $\gtrsim 1$ year (specifically we chose $\Delta t > 2\times 10^7$s).  We did not perform this step if the initial radius of the model was less than the radius of the \textsc{Kepler} model that was disrupted.  Whenever timestep exceeded $10^6$ s, we allowed fine rezoning of the outer layers to re-establish an atmosphere that can not be modelled in \textsc{Phantom}, as it does not have sufficient resolution.

 As the models were not in thermal equilibrium in \textsc{Kepler} after mapping, we estimated an approximate Kelvin-Helmholtz timescale for the models to settle into thermal equilibrium, 
\begin{equation}
    t_{\rm KH} = \frac{GM_*}{2R_* L_*}\;.
\end{equation}
using approximate main-sequence values for
\begin{subequations}
\begin{align}
    R_* &= R_\odot\left(\frac{M_*}{M_\odot}\right)^x,\quad\text{where}\; x={%
    \begin{cases}
        0.6 & \text{if } M_* > 1\,\mathrm{M}_\odot \\
        0.8 & \text{otherwise}
    \end{cases}} \\
    L_* &= L_\odot\left(\frac{M_*}{M_\odot}\right)^y,\quad\text{where}\;y={%
    \begin{cases}
        4 & \text{if } M_* > 1\,\mathrm{M}_\odot \\
        3 & \text{otherwise}
    \end{cases}}
\end{align} 
\end{subequations}

Figure~\ref{fig:hr_comparing_models} shows the evolution of a $1\,\mathrm{M}_\odot$, $3\,\mathrm{M}_\odot$, and $10\,\mathrm{M}_\odot$ ZAMS models that were mapped into \textsc{Phantom}, and then mapped back into \textsc{Kepler}.  The evolution is similar to the original \textsc{Kepler} models, hence, showing the accuracy of our mapping back procedure.  We note that for a few models, we had to ignore the outermost five bins to successfully map them into \textsc{Kepler} as otherwise, these models crashed. 
 These models were $1\,\mathrm{M}_\odot$ TAMS, $\beta=1.08$; $3\,\mathrm{M}_\odot$ ZAMS, $\beta=1.29$; $10\,\mathrm{M}_\odot$ ZAMS, $\beta=0.99$; $10\,\mathrm{M}_\odot$ MAMS, 
  $\beta=0.98$ and $\beta=1.95$; $10\,\mathrm{M}_\odot$ TAMS, $\beta=0.94$.  
We also compared the remnants with \textsc{Kepler} models of the same mass. 
% If the remnant had mass $< 1.3 M_\odot$, then we used a mixing length parameter of 0.82 and multiplier on OPAL opacities of 1.175, otherwise we used 1 for both these values. We also used a minimum temperature of 1000 K and if the column depth was not less than 1 $g/cm^2$, then we set minimum density to $10^{-12} g/cm^3$, but once it was less than 1 $g/cm^2$, we reset it to the default value of $0.0001 g/cm^3$. 
% Luminosity of remnant and original model plot

\section{Results}
\begin{figure*}
	% To include a figure from a file named example.*
	    % Allowable file formats are eps or ps if compiling using latex
	% or pdf, png, jpg if compiling using pdflatex

    \centering
	\includegraphics[width=\textwidth]{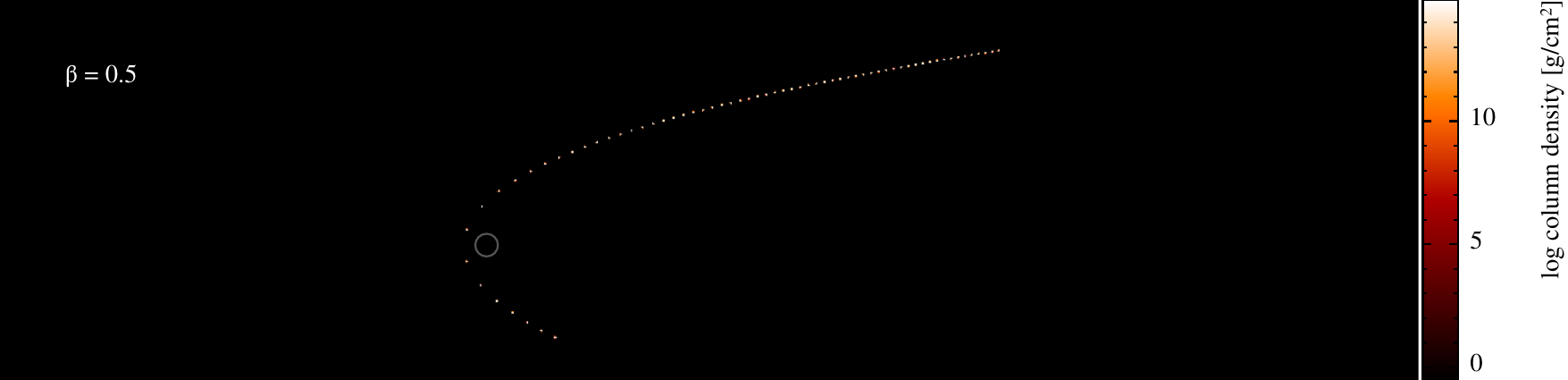}
	\includegraphics[width=\textwidth]{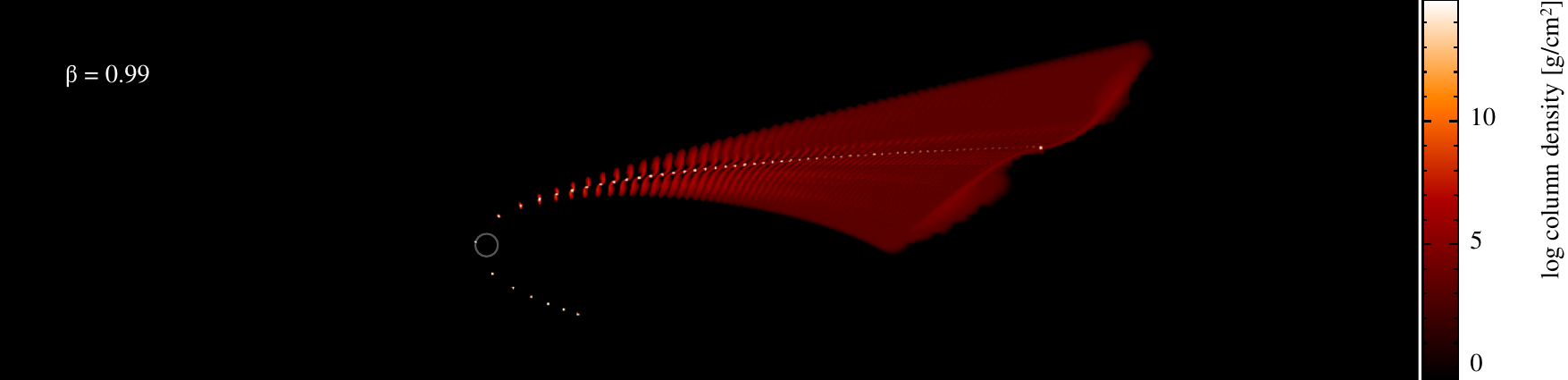}
	\includegraphics[width=\textwidth]{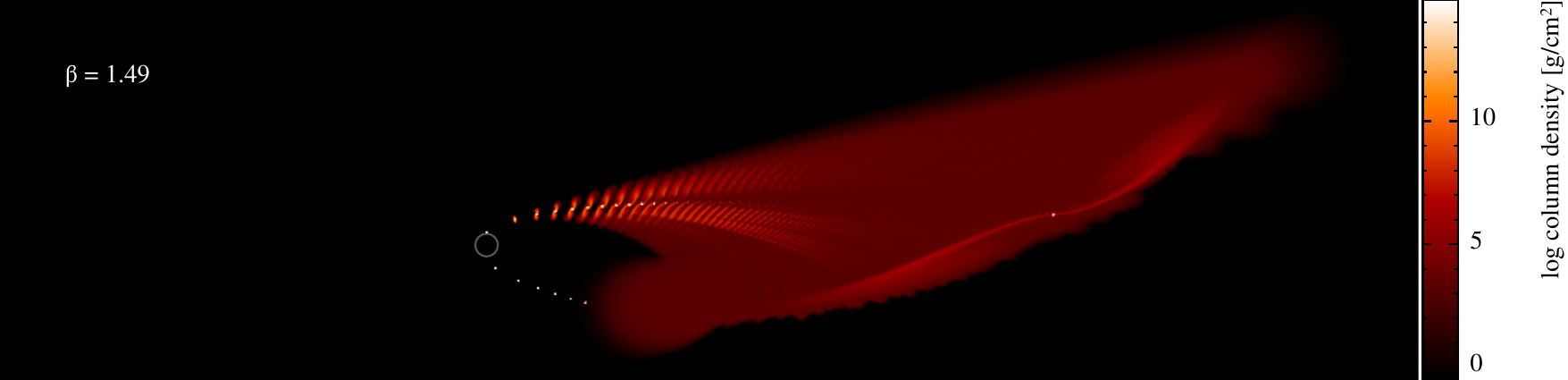}
	\includegraphics[width=\textwidth]{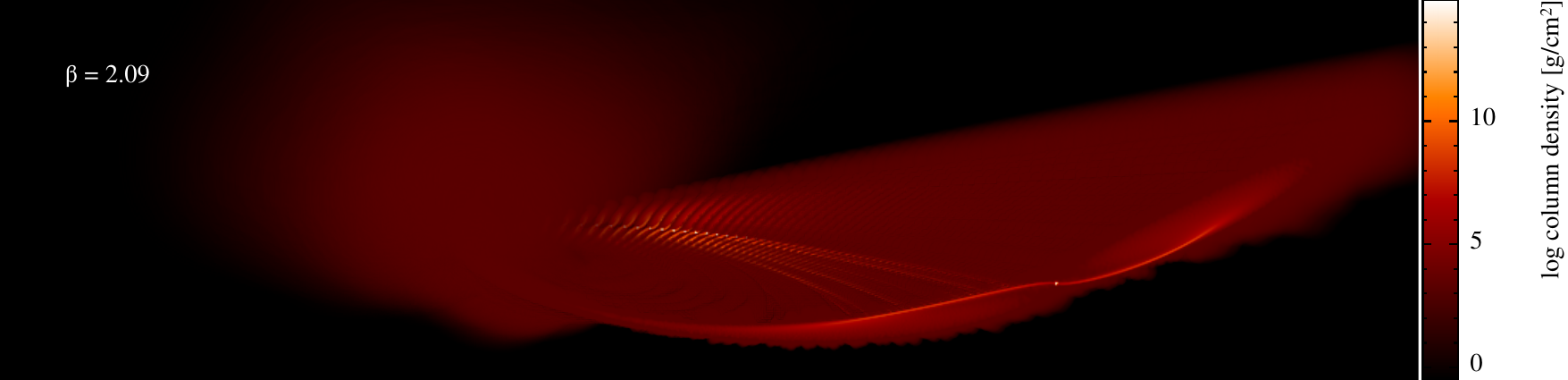}
	\includegraphics[width=\textwidth]{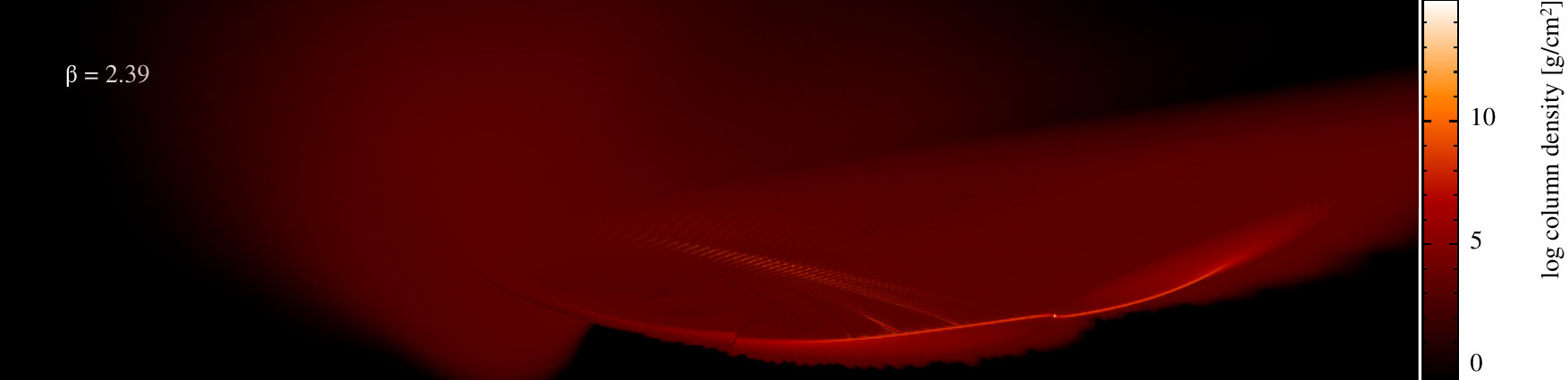}
    \caption{Snapshots of $3\, \mathrm{M}_\odot$ MAMS  model disrupted at different impact parameter ($\beta$; \textit{top} to \textit{bottom}), showing the column density perpendicular to the orbital plane.  The grey circle corresponds to the tidal radius.  All simulations were evolved for 8 days and each snapshot was at a $3\,\mathrm{h}$ interval.  The star loses more mass as it approaches the SMBH at a closer distance.  Streams are formed around the remnant which fall back onto the remnant.}
    \label{fig:phantom_disruption}
\end{figure*}

\label{sec:results_sec}
\subsection{What happens to the star?}
 
Figure~\ref{fig:phantom_disruption} shows snapshots of the \textsc{Phantom} simulations of the disruption of a $3\,\mathrm{M}_\odot$ MAMS model for a range of penetration factors, $\beta$ (\textit{top} to \textit{bottom}), showing the column density perpendicular to the orbital plane, with snapshots superimposed at every $3\, \mathrm{h}$ within each panel up to 8 days since the beginning of the simulation.  For an initial $\beta=0.8$ the star can be seen to have lost almost no mass, whereas for $\beta=2.39$, we can observe that the star loses about $80\%$ of its mass.  Post-pericentre passage, we see the formation of streams around the remnants.  Eventually, some of this material falls onto the SMBH.  Several authors have tried to determine the power-law fits of this accreting material onto the SMBH \citep{Guillochon2013,Goicovic2019,Goligothly2019,Ryu32020}.  In this paper, we focus on the evolution of the remnants themselves. 
 
Table~\ref{tab:models_used} lists the central density and the radius of the stars we disrupted in \textsc{Phantom}.  As a star evolves from ZAMS to TAMS, it becomes more dense and its radius increases.  This affects how long it can survive the encounter.  For very deep encounters, the star can be completely ripped apart, fall back, and eventually reassemble itself into a remnant.  We saw this behaviour for some of our closest encounters with the SMBH. An example is the $10\,\mathrm{M}_\odot$ MAMS, $\beta=2.15$ remnant of $0.24\,\mathrm{M}_\odot$.  The central density decreases for the first $14$ days, but afterwards, it increases and becomes constant as a remnant is formed (see Appendix~\ref{app:sectionrecollapse}).  This model has $r_\mathrm{p}~=~115.80~\mathrm{R}_\odot$ and exhibits complete disruption and then re-collapse.  We performed simulations for different stars where they approached the SMBH at even lower $r_\mathrm{p}$ values, but did not undergo complete disruption and then re-collapse.  \citet{Guillochon2013} and \citet{NixonC2022} had used Newtonian physics but also found that the streams can re-collapse to form a remnant for their polytropic models.  \citet{NixonC2022} argued that this is a result of gravitational instability in the stream.

\subsection{Stellar oscillations}
\begin{figure}
	\includegraphics[width=\columnwidth]{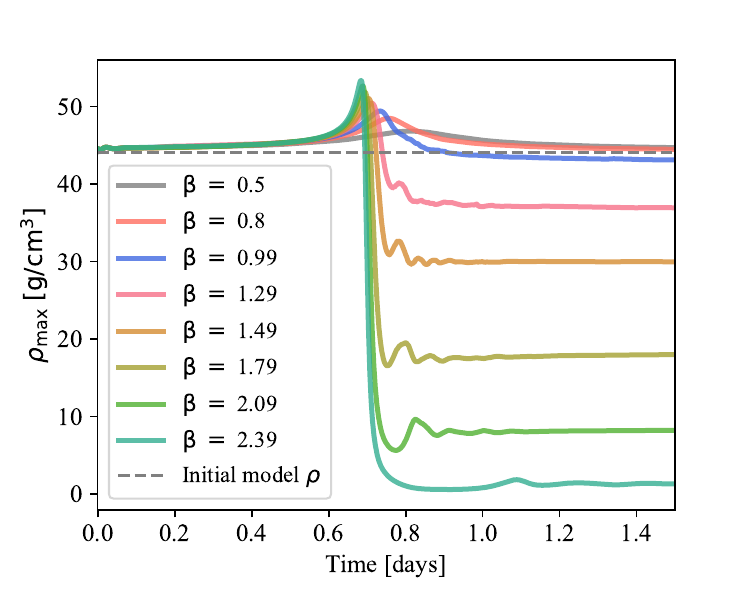}

    \caption{Central density as a function of time for $3\,\mathrm{M}_\odot$ MAMS models at different $\beta$ values since the beginning of the simulations.  The central density of the star reaches a peak at about $15$ hours after the beginning of our simulation and induces oscillations in the remnant star.  }
    \label{fig:max_den_time_oscill}
\end{figure}

Figure~\ref{fig:max_den_time_oscill} shows the central density of the $3\,\mathrm{M}_\odot$ MAMS remnants as a function of time for different penetration factors ($\beta$). We see that the density of the remnant starts to increase as it approaches the pericentre, becoming maximum around the pericentre, and then decreasing as the star moves away from the SMBH.  It eventually settles to a constant value.  Even though the $\beta=0.5$ model does not lose any mass, the star experiences a slight compression which eventually dissipates. 

We find that the encounter induces oscillations in the remnant star.  The period of this ringing is close to that of the fundamental mode of the star, $(t_{\rm osc} = \sqrt{3\pi/(32G\bar{\rho})}\,)$.  For example, we looked at the $\beta=1.79$ case for $3\,\mathrm{M}_\odot$ MAMS star.  We considered the material bound within $2\,\mathrm{R}_\odot$
as part of the remnant.  This resulted in a oscillation time of $0.04$ days, whereas $t_{\rm osc} \sim 0.07$ days.  We observe a similar trend in the central temperature as a function of time. 

\begin{figure}
    \centering
    \includegraphics[width=\columnwidth]{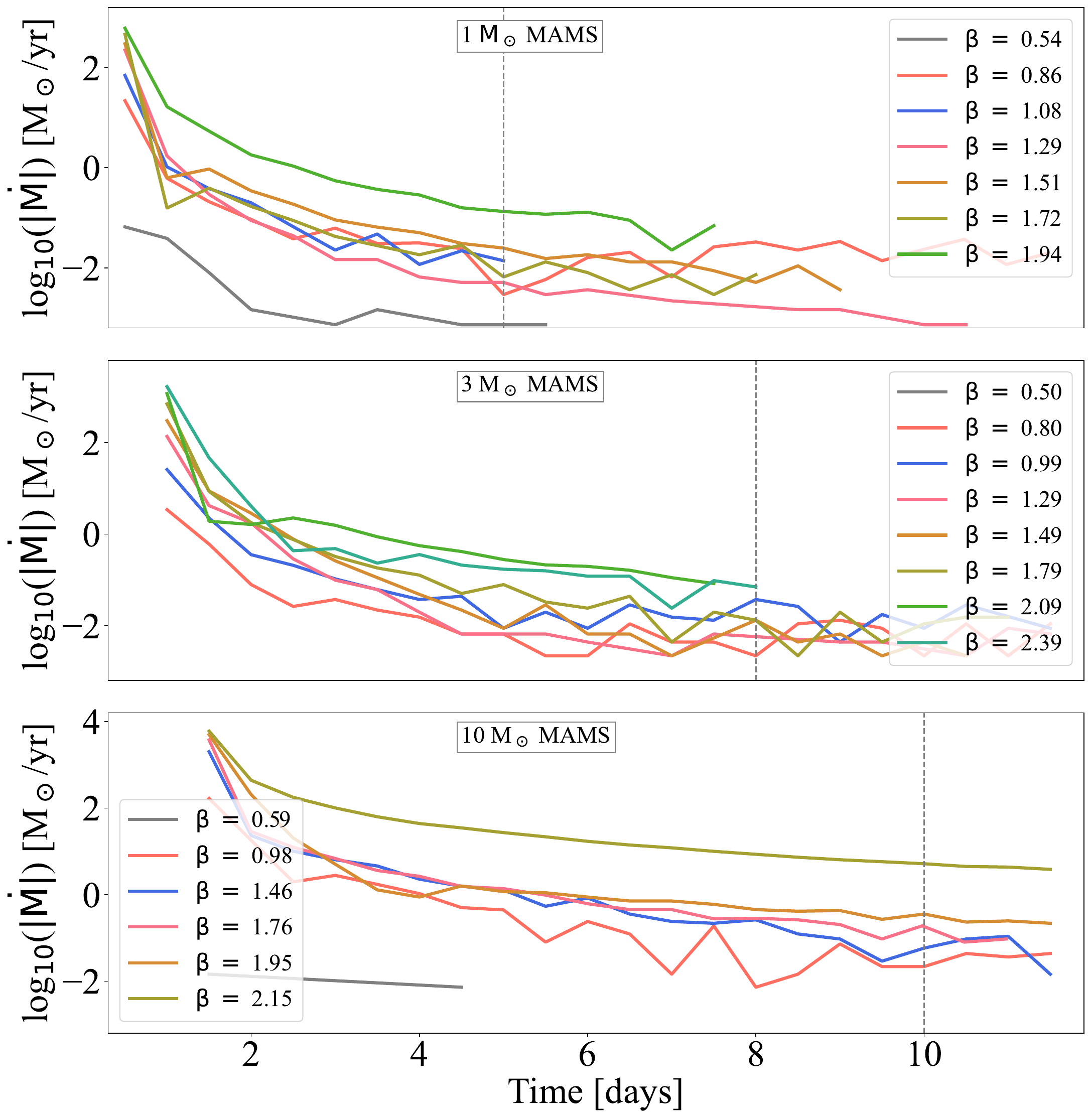}  
    \caption{Absolute accretion rate onto the remnant as a function of time since the beginning of the simulation, calculated by considering the difference in mass of the remnant every $12$ hour.  The grey line represents the time at which we perform analysis for our models.  We selected times of $5$ days, $8$ days, and $10$ days for $1\,\mathrm{M}_\odot$, $3\,\mathrm{M}_\odot$, and $10\,\mathrm{M}_\odot$ models, respectively for our analysis.  For $10\,\mathrm{M}_\odot$, $\beta = 2.15$ model, we chose the snapshot at $60$ days.  }
    \label{fig:accretion_rate}
\end{figure}

\subsection{Accretion onto the remnant}
Figure~\ref{fig:accretion_rate} shows the mass accretion rate onto the remnant as a function of time, determined by calculating the difference in the mass of the remnant over $12$ hours intervals.  For most of the encounters shown in Figure~\ref{fig:accretion_rate}, the mass accretion rate can be seen to become negligible after several days.  However, we observed a persistent, albeit small, stream of material continuing to accrete over longer timescales.  This stream moves away from the remnant as the simulation progresses. 

For example, we found some material still accreting onto the remnant at $100$ days for $\beta=1.79$ and a $3\,\mathrm{M}_\odot$ MAMS model.

\subsection{Mass of remnants}

Table~\ref{tab:models_run} lists the remnant masses for each model.  As anticipated, TAMS model stars can survive higher $\beta$ encounters compared with the ZAMS models, owing to their more centrally concentrated structure.  Recall that for these stars, the same $\beta$ corresponds to more distant encounters compared to ZAMS models because these stars have initially lower average densities.  For $\beta \lesssim 1$, stars exhibit varying degrees of mass losses, ranging from cases with no mass loss ($\beta \lesssim 0.5$) to others with losses $\leq 10\%$.  

\begin{figure}
    \centering
    \includegraphics[width=\columnwidth]{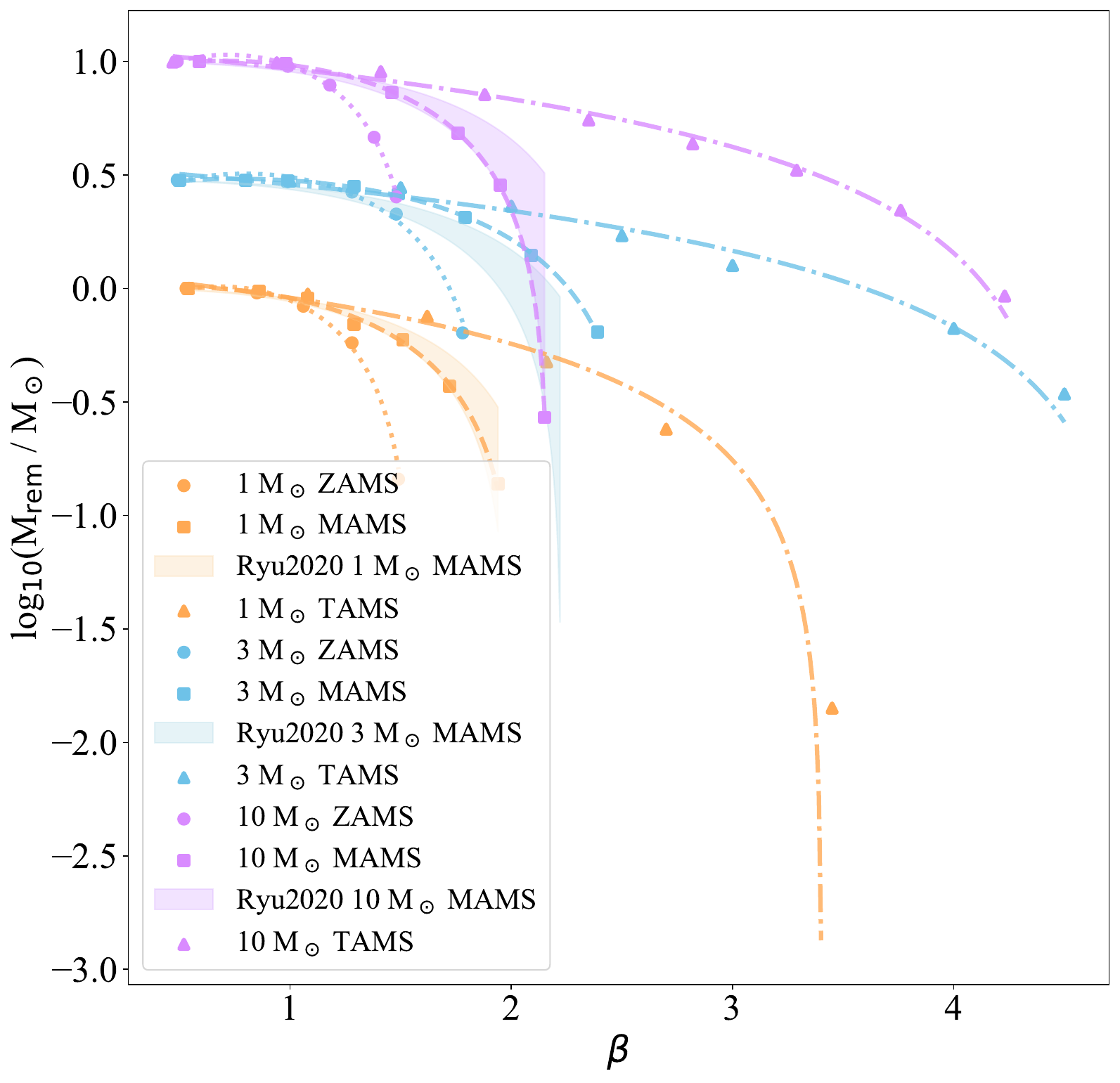}
    \caption{Mass of remnant as a function of penetration factor, $\beta$.  The lines connecting the points represent the fitting functions that were determined for each model.  We also plot the remnant masses based on the Newtonian functional form given by \citet{Ryu12020} as shaded regions for their MAMS models.  More evolved stars (MAMS, TAMS) can survive the encounter closer to the SMBH compared to the ZAMS because as stars evolve, the central density and radius increase.  }
    \label{fig:remnant_mass_vs_beta}
\end{figure}

Figure~\ref{fig:remnant_mass_vs_beta} shows the mass of the remnant as a function of the penetration factor, $\beta$. The remnant mass is defined as described in Section~\ref{sec:method_rem}.  Additionally, we plot polynomial fits given by
\begin{equation}
\label{eq:rem_mass_parm}
    \frac{M_{\mathrm{rem}}}{M_\star} = a\,\beta^2 + b\,\beta +c ,
\end{equation}
where $a,b,c$ are the parameters determined by fitting, as lines connecting our data points.  Table~\ref{tab:fit_params} gives the fitted parameters as well as the root mean squared error in the fit.

\begin{table}
\centering
 \caption{Column 1 gives the mass of the stellar model.  Column 2 lists the stage of evolution.  Columns 3, 4, and 5 list the parameters of the polynomial fit.  Column 6 gives the root mean squared error.   }
 \label{tab:fit_params}
 \begin{tabular}{cccccc}
  \hline
   \hline 
  Mass & Stage & $a$ & $b$ & $c$ & Root mean   \\
  (M$_\odot$) &  &  &  & &squared error \\
  % \hline
  \hline 
  1 & ZAMS & -1.31 & 1.78 & 0.41 &  0.02 \\
  1 & MAMS & -0.43 & 0.43 & 0.90 & 0.03\\
  1 & TAMS & -0.03 & -0.26 & 1.19  & 0.05\\
  3 & ZAMS & -0.86 & 1.37 & 0.52 & 0.1\\
  3 & MAMS & -0.31 & 0.46 & 0.83 & 0.03\\
  3 & TAMS & -0.01 & -0.2 & 1.16 & 0.15\\
  10 & ZAMS & -1.39 & 1.99 & 0.36 & 0.1 \\
  10 & MAMS & -0.47 & 0.65 & 0.78 & 0.05\\
  10 & TAMS & -0.01 & -0.22 & 1.16 & 0.39\\
  \hline
   \hline 
 \end{tabular}
\end{table}

\citet{Ryu12020} determined a functional form for mass of remnants in the Newtonian limit as $M_\mathrm{rem}/M_\star = 1 - (r_\mathrm{p}/\mathcal{R_\mathrm{t}})^{-3}$ where $\mathcal{R_\mathrm{t}}$ is the \emph{physical} tidal radius --- defined as the maximum distance at which star would experience full disruption.  Since $r_\mathrm{t}$ does not account for the internal structure of the star, it can survive the disruption at distances $r_\mathrm{p} < r_\mathrm{t}$, aligning with our findings.  \citet{Ryu2020b} provides $\mathcal{R_\mathrm{t}}/r_\mathrm{g}$ values of $22.5\pm1.2$, $33.9\pm2.0$, and $52.1\pm3.1$ for their $1\,\mathrm{M}_\odot$, $3\,\mathrm{M}_\odot$, and $10\,\mathrm{M}_\odot$ MAMS models, where $r_\mathrm{g}=2GM_\bullet/c^2$. 
 They also list the $r_\mathrm{t}/r_\mathrm{g}$ as $47.5$, $79.8$, and $123$, respectively. Utilising these values, we calculated the $M_\mathrm{rem}$ for a range of $\beta$ values based on the $\beta$ of our models. The results are shown as shaded regions in Figure~\ref{fig:remnant_mass_vs_beta}.  Notably, our $1\,\mathrm{M}_\odot$ and $10\,\mathrm{M}_\odot$ MAMS models fall within these shaded regions, but our $3\, \mathrm{M}_\odot$ remnant masses surpass the expected range.  We also observed that the remnant mass calculated by this functional formula is negative for higher $\beta$ values, even in instances where remnants are still present for our simulations. Such an example is $3\,\mathrm{M}_\odot$ MAMS disrupted for a $\beta = 2.39$.  This inconsistency may stem from differences in the internal structures of our models.

\subsection{Critical $\beta$ ($\beta_\mathrm{c}$)}
\label{sec:critical_beta_sec}
We determined the critical $\beta$, beyond which our stellar models would experience complete disruption ($\beta_\mathrm{c}$), using the polynomial fits.  The roots of these functions were determined, corresponding to $\beta$ values where the remnant mass becomes zero.  Values and uncertainties of $\beta_\mathrm{c}$ are listed in Table~\ref{tab:models_used}. These uncertainties correspond to the coarseness of the simulations run.  The fitting function determined a $\beta_\mathrm{c} = 3.4\pm0.5$ for the $1\,\mathrm{M}_\odot$ TAMS model. We found that a remnant can form , $\beta=3.45$ albeit of only $0.01\,\mathrm{M}_\odot$.

 \citet{Lawsmith2020} and \citet{Ryu2020b} argued that $\beta_\mathrm{c}$ is related to the ratio of critical density to the mean density.  
 
\begin{figure}
    \centering
    \includegraphics[width=\columnwidth]{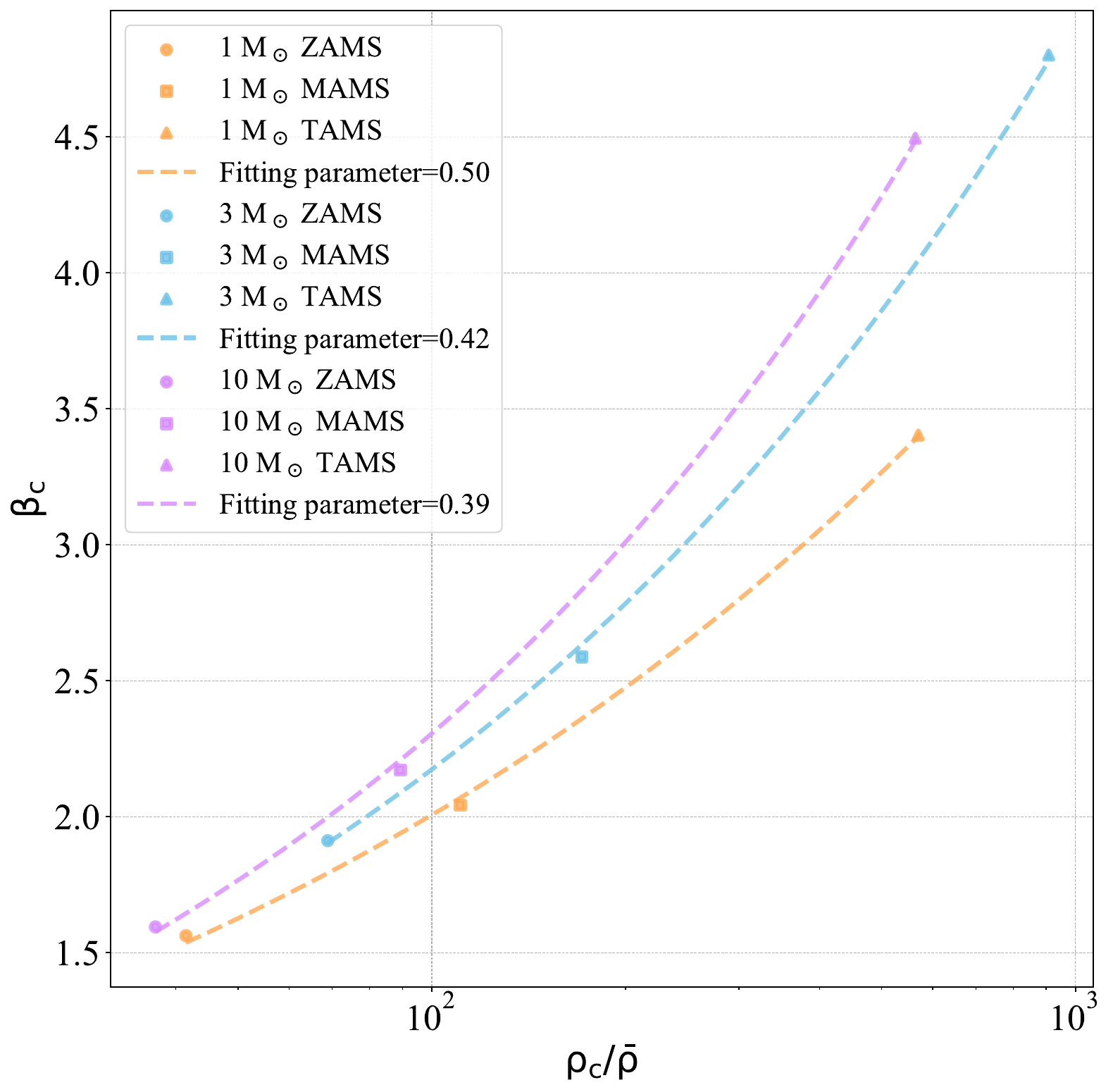}
\caption{Critical $\beta$ or the point where the remnant mass is zero ($\beta_\mathrm{c}$) plotted as a function of $\rho_\mathrm{c}/\bar{\rho}$.  Our fit functions are given in Eq.~\ref{eq:fit_form}. $1\,\mathrm{M}_\odot$ TAMS and $10\,\mathrm{M} _\odot$ TAMS have similar $\rho_\mathrm{c}/\bar{\rho}$. 
 Hence, we fit individual functions for each star.}
    \label{fig:beta_vs_den}
\end{figure}

 Figure~\ref{fig:beta_vs_den} shows the critical $\beta$ as a function of the ratio of central density to the mean density of the initial stellar model.  We conducted fits between $\beta_\mathrm{c}$ and $(\rho_c/\bar{\rho})$ and determined the following fit functions: 
\begin{subequations}
\label{eq:fit_form}
\begin{align}
    \beta_c &= 0.50\times(\rho_c/\bar{\rho})^{1/3.3} \quad \text{for $1\,\mathrm{M}_\odot$,} \\
     \beta_c &= 0.42\times(\rho_c/\bar{\rho})^{1/2.8} \quad \text{for $3\,\mathrm{M}_\odot$,} \\
     \beta_c &= 0.39\times(\rho_c/\bar{\rho})^{1/2.6} \quad \text{for $10\,\mathrm{M}_\odot$.}
\end{align}
\end{subequations}
 

\subsection{Properties of the remnants}

\begin{figure*}
    \centering
    \includegraphics[width=\textwidth]{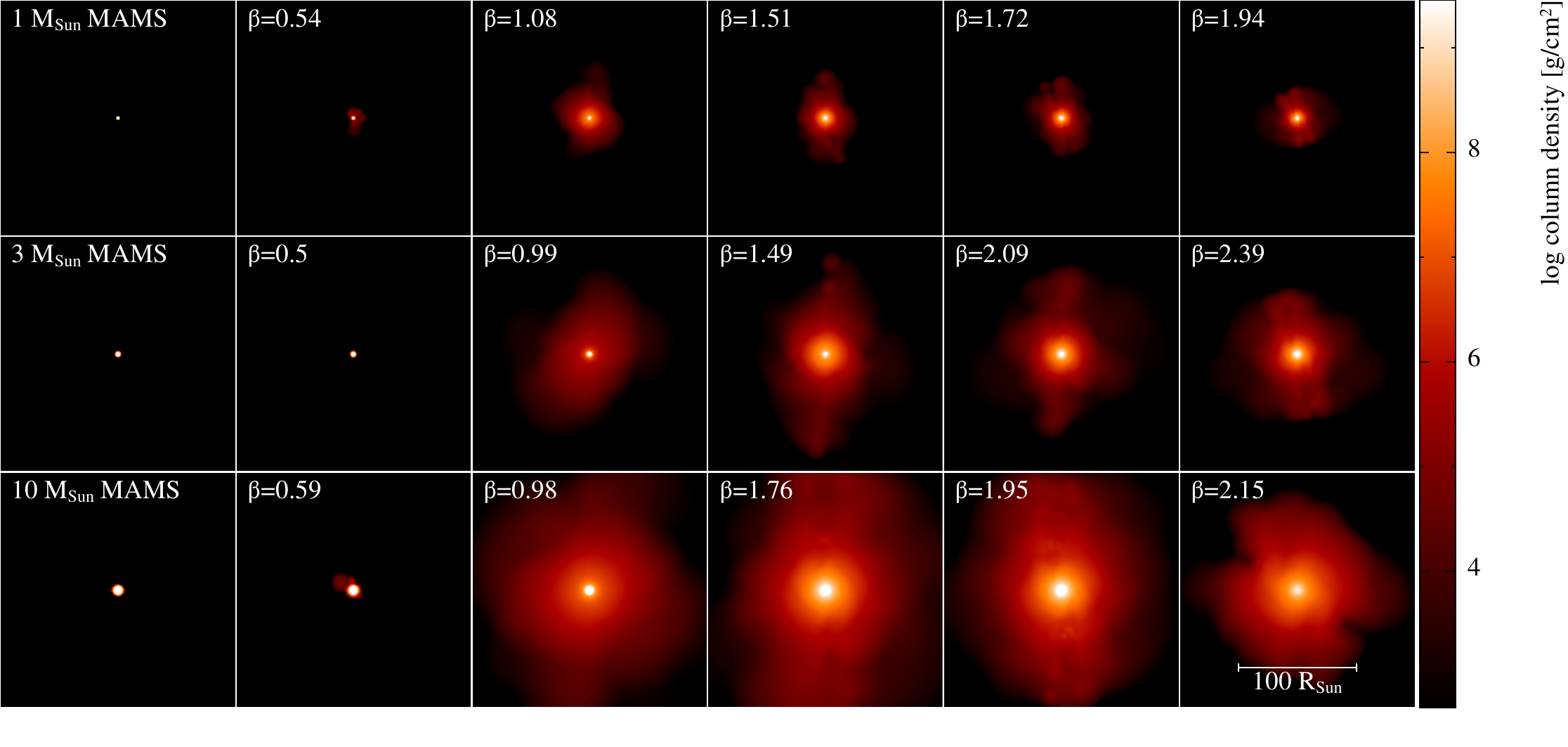}
    \caption{Remnants of $1\,\mathrm{M}_\odot$, $3\,\mathrm{M}_\odot$, and $10\,\mathrm{M}_\odot$ MAMS models at $4.8$ days, $7.3$ days, and $9$ days post-pericentre, respectively (\textit{top} to \textit{bottom}).  Except, the $10\;\mathrm{M}_\odot$, $\beta=2.15$ is analysed at $59$ days post-pericentre.  We show column density perpendicular to the orbital plane, showing only material identified as belonging to the remnant (i.e. bound and not belonging to the streams).  Each panel is $200\,\mathrm{R}_\odot \times 200\,\mathrm{R}_\odot$.  Each row shows the models at different penetration factors, with the first column showing the models before the disruption.  The radius of the remnants increases as the star approaches closer to the SMBH.  The remnants bear a striking similarity to G-objects in the galactic centre \citep{Ciurlo2020}. }
    \label{fig:rendered_models_new}
\end{figure*}

Figure~\ref{fig:rendered_models_new} shows the column density perpendicular to the orbital plane showing only the material identified as the stellar remnants, as per Section~\ref{sec:method_rem}, for $1\,\mathrm{M}_\odot$, $3\,\mathrm{M}_\odot$, and $10\,\mathrm{M}_\odot$ MAMS models (\textit{top} to \textit{bottom}) for different initial $\beta$ values (\textit{left} to \textit{right}).  The models are shown at $4.8$ days, $7.3$ days, and $9$ days post-pericentre (i.e., $5$ days, $8$ days, and $10$ days since the start of the simulation).  The first column shows the initial undisrupted star.  The remnants exhibit dense cores ($\sim$ radius of the initial star) with fluffy outer envelopes extending to $\sim10$--$100\,\mathrm{R}_\odot$ for the most penetrating encounters, a result of material stripping leading to expanded radii.  The radius of the remnant also increases with increasing $\beta$.  The remnant's radius increases with higher $\beta$, e.g., the $1\,\mathrm{M}_\odot$ model's remnant is nearly $10$ times larger than its initial radius for $\beta > 1$.  For encounters with pericentre more distant than the tidal radius (e.g., $\beta = 0.54$) we find only a slight increase in the radius of the remnant compared to the initial star, which is due to lower stripping of the star.   

\begin{figure*}
\centering
\includegraphics[width=\textwidth]{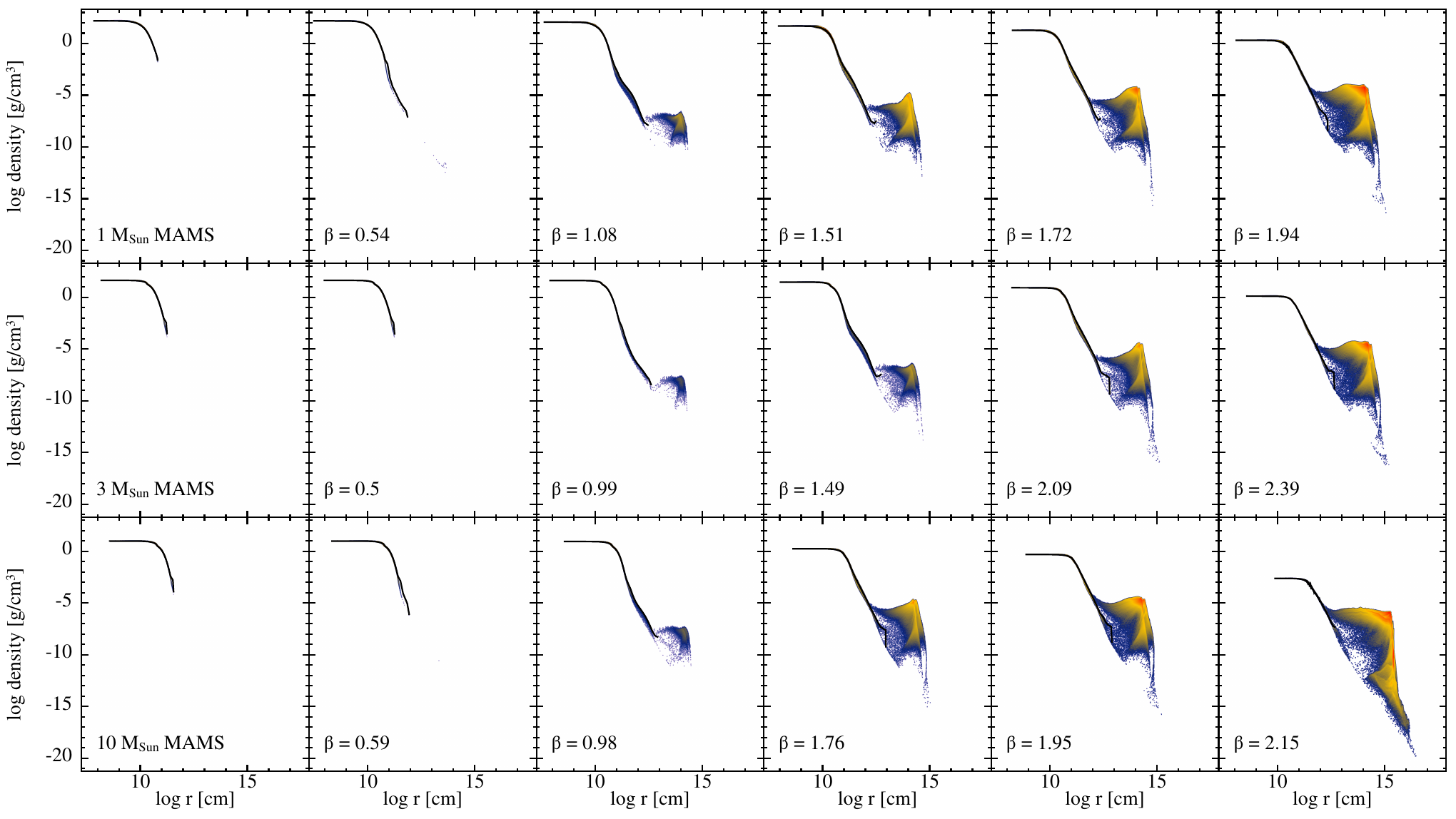}
\caption{Density as a function of radius for the same models shown in Figure~\ref{fig:rendered_models_new}.  We have also plotted the radially-binned density as a black solid line.  The density decreases as the star approaches the SMBH at a closer distance (higher $\beta$, \textit{left} to \textit{right}).  }
\label{fig:Binned_density}
\end{figure*}

Figure~\ref{fig:Binned_density} shows density as a function of radius for the same set of models shown in Figure~\ref{fig:rendered_models_new}.  The first column presents a comparison of the density distribution for the models mapped into \textsc{Phantom} without tidal disruption.  The binning algorithm is observed to maintain the expected density profile.  The central density of the remnant can be seen to decrease as the star passes closer to the SMBH (increasing $\beta$; \textit{left} to \textit{right}).  For instance, the remnant from the $\beta = 1.94$ simulation of $1\,\mathrm{M}_\odot$ exhibits a post-disruption central density of $2.02\,\mathrm{g}\,\mathrm{cm}^{-3}$, whereas for $\beta = 0.54$, the central density is $150.84\,\mathrm{g}\,\mathrm{cm}^{-3}$.  In very weak encounters like $\beta=1.08$, the remnant loses some material ($\sim 3\%$), but the central density remains similar to the original model.  We have also plotted the radially-binned density (Section~\ref{sec:binning_algo}) on the same plot for comparison, showing that our remnants are generally spherical except in the outermost layers where material continues to accrete onto the remnant. 

\subsection{Rotational properties}
\label{sec:rotational_properties}
\begin{figure*}
    \includegraphics[width=\textwidth]{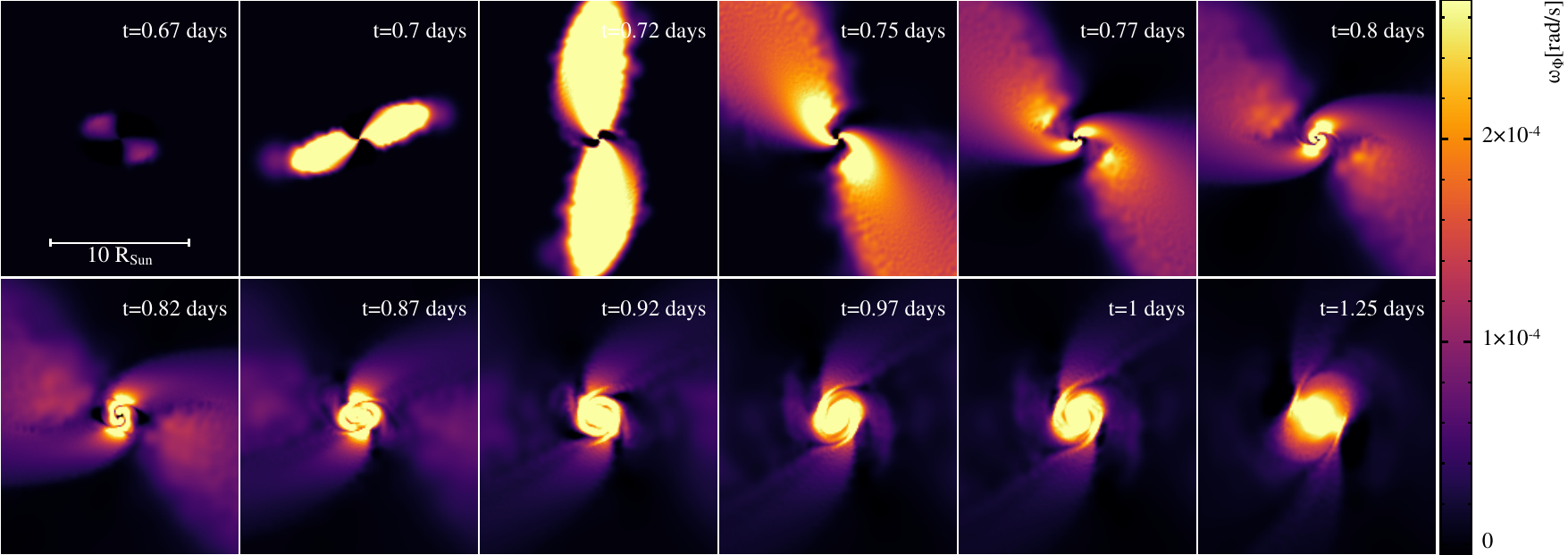}
    
    \caption{Time evolution of angular velocity in cross-section slice in the orbital plane, for $3\,\mathrm{M}_\odot$ MAMS, $\beta=1.79$ model since the beginning of the simulation.  Each panel is $20\,\mathrm{R}_\odot \times 20\,\mathrm{R}_\odot$.  Vortices form at about $16 \,\mathrm{h}$ (first panel), which results in angular momentum transfer.  This results in the central region rotating faster than the outer layers of the star at about $24\,\mathrm{h}$.}
    \label{fig:rotaion_mixing}
\end{figure*}

\begin{figure*}
    \centering
    \includegraphics[width=\textwidth]{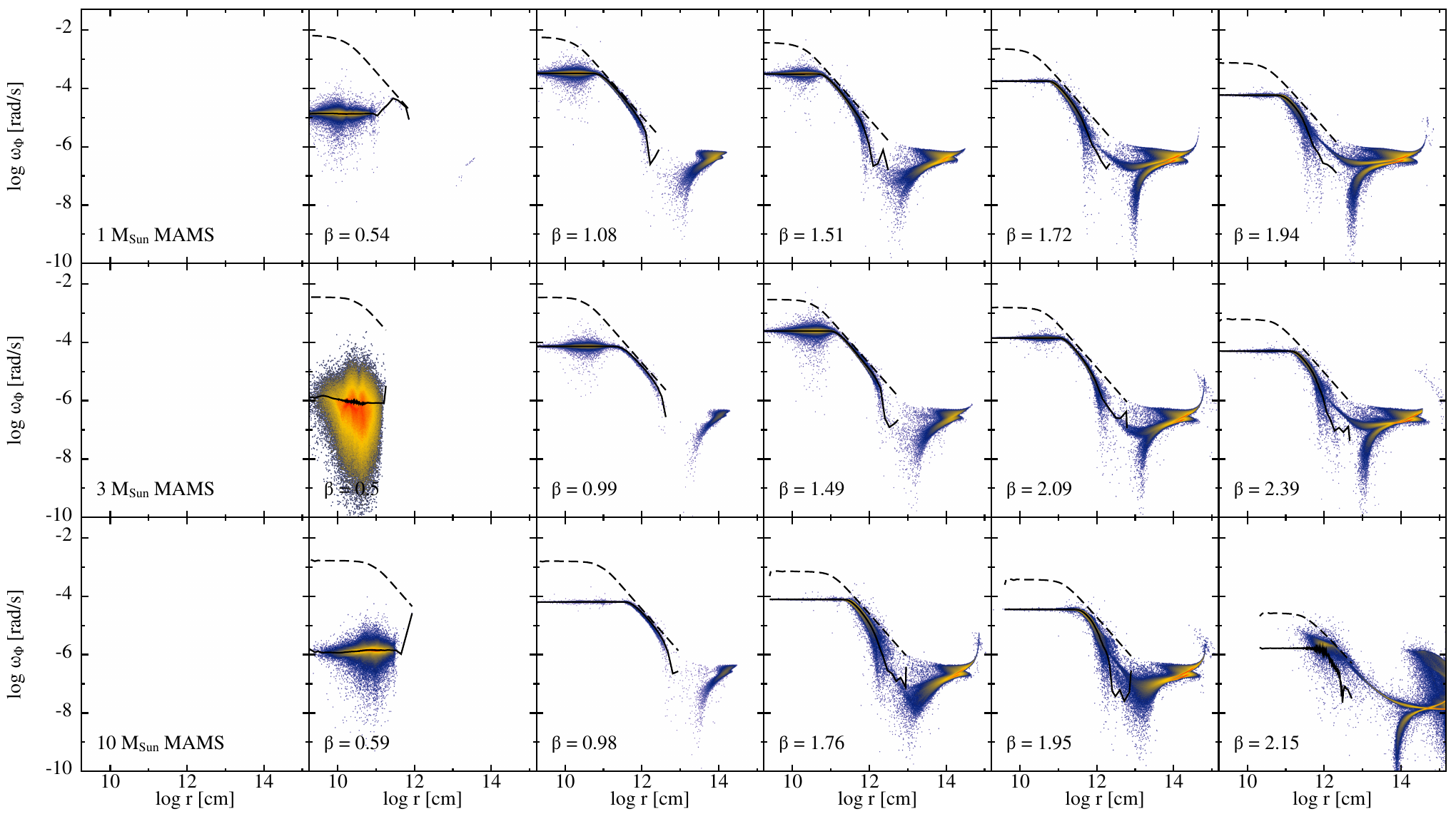}
    \caption{$\omega_\phi$ as function of radius for remnants of $1,\mathrm{M}_\odot$, $3,\mathrm{M}_\odot$, and $10\,\mathrm{M}_\odot$ MAMS models at $4.8$ days, $7.3$ days, and $9$ days since disruption, respectively (\textit{top} to \textit{bottom}).  Each row shows the models at different penetration factors, with the first column showing the models before the disruption.  The black line shows the radially-binned angular velocity and the dashed line represents the break-up velocity.  Initially, the stars have no angular velocity.  Following the disruption, the remnant's region with a radius close to the initial star's radius rotates rigidly, while the outer layers exhibit differential rotation.  }
    \label{fig:omega_cylindrical}
\end{figure*}
%We performed visual assessment of angular velocity in both cylindrical and spherical coordinates to determine if the star rotates on cylinders or shells, but could not confirm this behaviour. Hence. we have used cylindrical coordinates for our analysis in this paper.  

When a star is tidally disrupted by a SMBH, one might expect that the outer layers rotate faster than the core due to the stronger impact of tides.  This, however, is not what happens in our simulations.  Figure~\ref{fig:rotaion_mixing} shows the time evolution of the angular velocity ($\omega_\mathrm{\phi}$), computed in cylindrical coordinates in a cross-section slice in the orbital plane for our $3\, \mathrm{M}_\odot$ MAMS model for a penetration factor $\beta=1.79$.  At $16\,\mathrm{h}$ before pericentre (\textit{top} left panel), we see the formation of two distinct regions in the radially stretched star, with one region rotating in the opposite direction.  These vortices transfer angular momentum from outer regions to the central region of the star, resulting in a remnant rotating faster in the central region ($\sim$ radius of the initial star) while the fluffy outer layers rotating slower ($t=24\,\mathrm{h}$).  The time at which these vortices disappear varies by model, being longer for closer encounters.  \citet{Goicovic2019} found something similar but their vortices were persistent structures, and were still present at $42\,\mathrm{h}$.

Figure~\ref{fig:omega_cylindrical} shows the angular velocity as a function of spherical radius.  Initially, our models exhibited no rotation, as shown in the first column.  We have also plotted the binned angular velocity (solid lines) and break-up velocity (dashed lines) given by   
\begin{equation}
    \omega_{\mathrm{breakup}} (R) \;=\;\sqrt{\frac{GM_\mathrm{enc}(R)}{R^3}}\;,
\end{equation}
for comparison.  $M_\mathrm{enc}$ is the enclosed mass at radius $R$.  For disruptions where the star loses almost no mass (e.g., $\beta = 0.54$ model for $1\,\mathrm{M}_\odot$ MAMS; second panel from left in the \textit{top} row), the remnant is rigidly rotating with a few outer zones having higher rotation.  None of the regions rotating rigidly have rotation close to the break-up velocity.  For $\beta \gtrsim 1$ (rightmost panels), the differentially rotating layers are close to the break-up velocities.  In summary, models that lose $\geq 40\%$ of their initial mass result in slower-rotating remnants compared to models that lose less mass.

\begin{figure*}
    \centering
    \includegraphics[width=\textwidth]{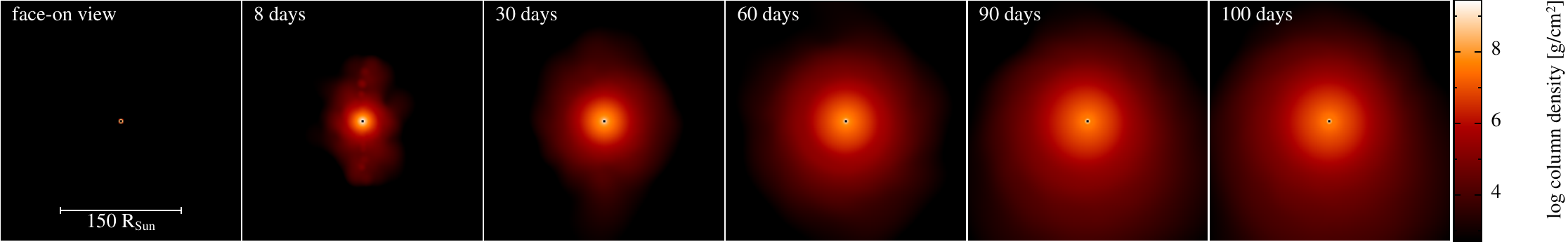}
    \includegraphics[width=\textwidth]{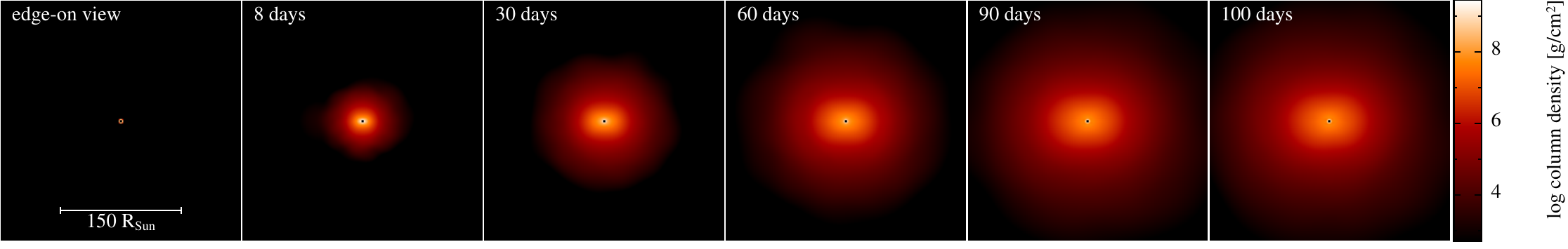}
    \caption{$3\,\mathrm{M}_\odot$, $\beta=1.79$ model remnant at different times (\textit{across}).  Each panel is $300\,\mathrm{R}_\odot \times 300\,\mathrm{R}_\odot$.  We show the face-on (\textit{top}) and edge-on (\textit{bottom}) view.  As time increases, the size of the remnant increases too.  There is the formation of differentially rotating circum-stellar envelope which is bound to the rigidly rotating central region.}
    \label{fig:disks1}
\end{figure*}

 Figure~\ref{fig:disks1} shows the formation of a circum-stellar envelope in the case of $3\,\mathrm{M}_\odot$ MAMS, $\beta=1.79$ simulation. The \textit{top} panel shows what is identified as a remnant based on our energy condition in the face-on view ($x$--$y$ plane), while the \textit{bottom} panel shows the edge-on view ($z$--$x$ plane) over $100$ days.  As time progresses, the outer layers of the remnant expand and result in the formation of an envelope that is bound to the remnant. 
 
\begin{figure}
    
    \includegraphics[width=\columnwidth]{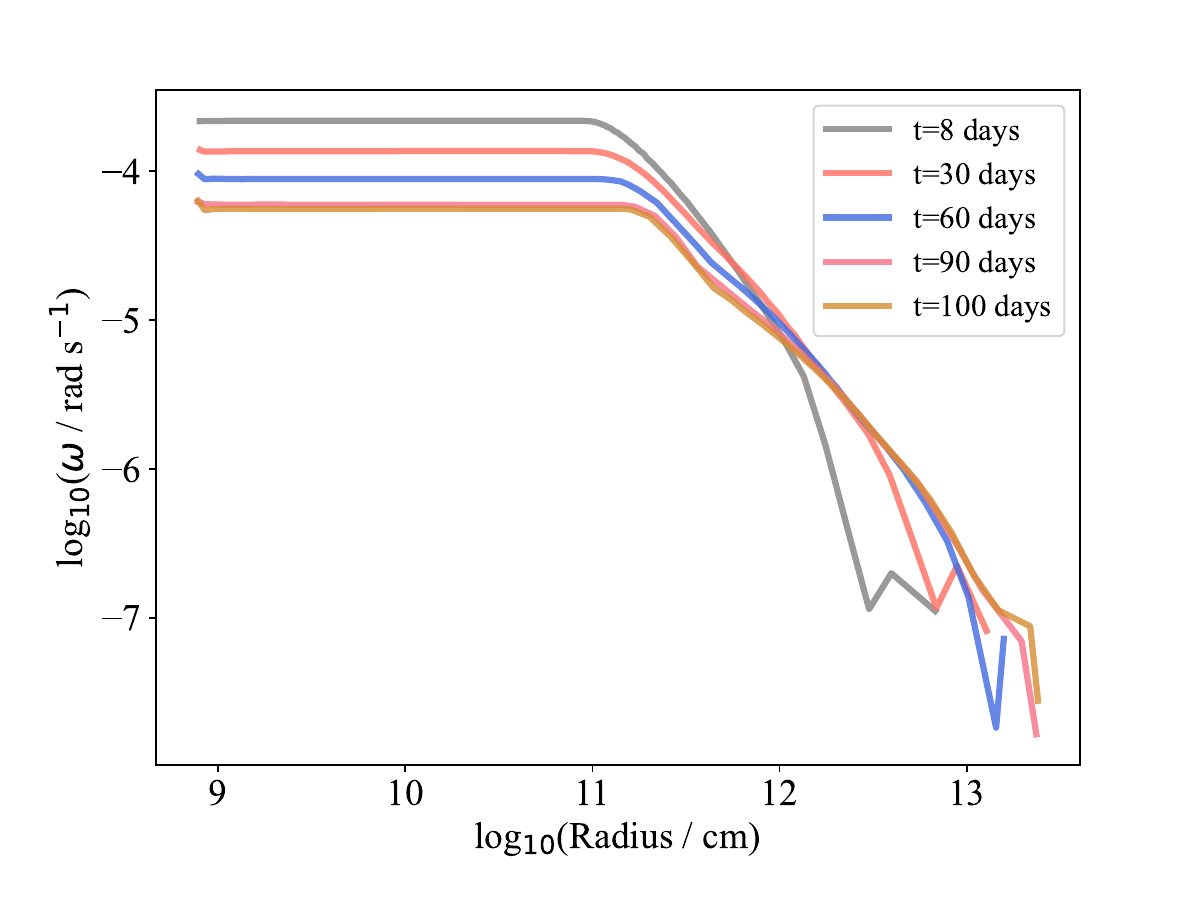}
    
    \caption{Binned angular velocity of remnant of $3\,\mathrm{M}_\odot$, $\beta = 1.79$ at different times.  As time increases, the rigid rotating layers rotate slower, while the differentially rotating region rotates faster.  This could be the result of angular momentum redistribution in the star.}
    \label{fig:angular_vel_wtr_time}
\end{figure}

Figure~\ref{fig:angular_vel_wtr_time} shows the evolution of angular velocity with time.  We found that as time progresses, the region that was rotating rigidly starts to rotate slower, and the differentially rotating zones, rotate faster.  This could be due to angular momentum transfer from the core to the outer region, and as the material is still falling back, it could also provide angular momentum to the outer layers.  This could affect the evolution of the stars, but as we did not implement angular velocity in our mapped back models, we can not quantify the possible difference.

\subsection{Escape velocity of remnants}
Table~\ref{tab:all_sims_data} lists the initial and final escape velocities derived from our simulations.  For the initial escape velocity, we followed the trajectory of a single point mass using a code that solves the geodesic equations \citep{Liptai2019}.  We started these simulations at $\sim 10\,r_\mathrm{t}$.  We progressed these simulations until the geodesic reached a distance $> 60\,r_\mathrm{t}$ for $1\,\mathrm{M}_\odot$, $3\,\mathrm{M}_\odot$, and $10\,\mathrm{M}_\odot$ models.  We calculated the velocity by determining \emph{Newtonian} kinetic and potential specific energy ($E = E_\mathrm{kinetic} + E_\mathrm{potential}$) using $v=\sqrt{2 E}$.  This approach remains valid at this stage as geodesics are sufficiently far away from the black hole, rendering the GR corrections negligible.

To determine the final escape velocity of the remnant, we initially calculated the centre of mass position and velocity relative to the SMBH at $5$ days, $8$ days, and $10$ days since the start of the simulations. 

As the encounter strength increases, the star's orbit undergoes more extreme changes.  We notice that TAMS models slow down while the ZAMS and the MAMS models tend to speed up.  For the same value of $\beta$ the more evolved TAMS stars have a larger periapsis distance which decreases the effects of general relativity.  Whether the post-encounter remnant is faster or slower than the incoming star, however, may depend on the detail of the tidal interaction, which could be affected by the interaction of the dynamical tide with oscillation resonances excited within the star \citep{Zahn1975,Mardling2001}.

We also found that a few remnants formed by disrupting $10\,\mathrm{M}_\odot$ TAMS model had negative energy.  We calculated the semi-major axis and eccentricity of these orbits by using $a=-G  (M_\bullet + M_\mathrm{rem})/(2E)$ and $e=1 - r_\mathrm{p}/a$.  The results are given in Table~\ref{tab:all_sims_data} and Table~\ref{tab:ecc_a_main}. 

We found that the velocity increases with time.  Hence, there is some variability in the presented results.  For example, for $3\,\mathrm{M}_\odot$, $\beta=4.5$ scenario, the remnant escape velocity is $384\, \mathrm{km}\,\mathrm{s}^{-1}$ at $8$ days, but at $20$ days, the velocity is $425\, \mathrm{km}\,\mathrm{s}^{-1}$.  The remnant mass and radius continue to change slightly with time, impacting the centre of mass calculation and the remnant's velocity.

\subsection{Composition mixing during disruption events}
\begin{figure*}

 \includegraphics[width=\textwidth]{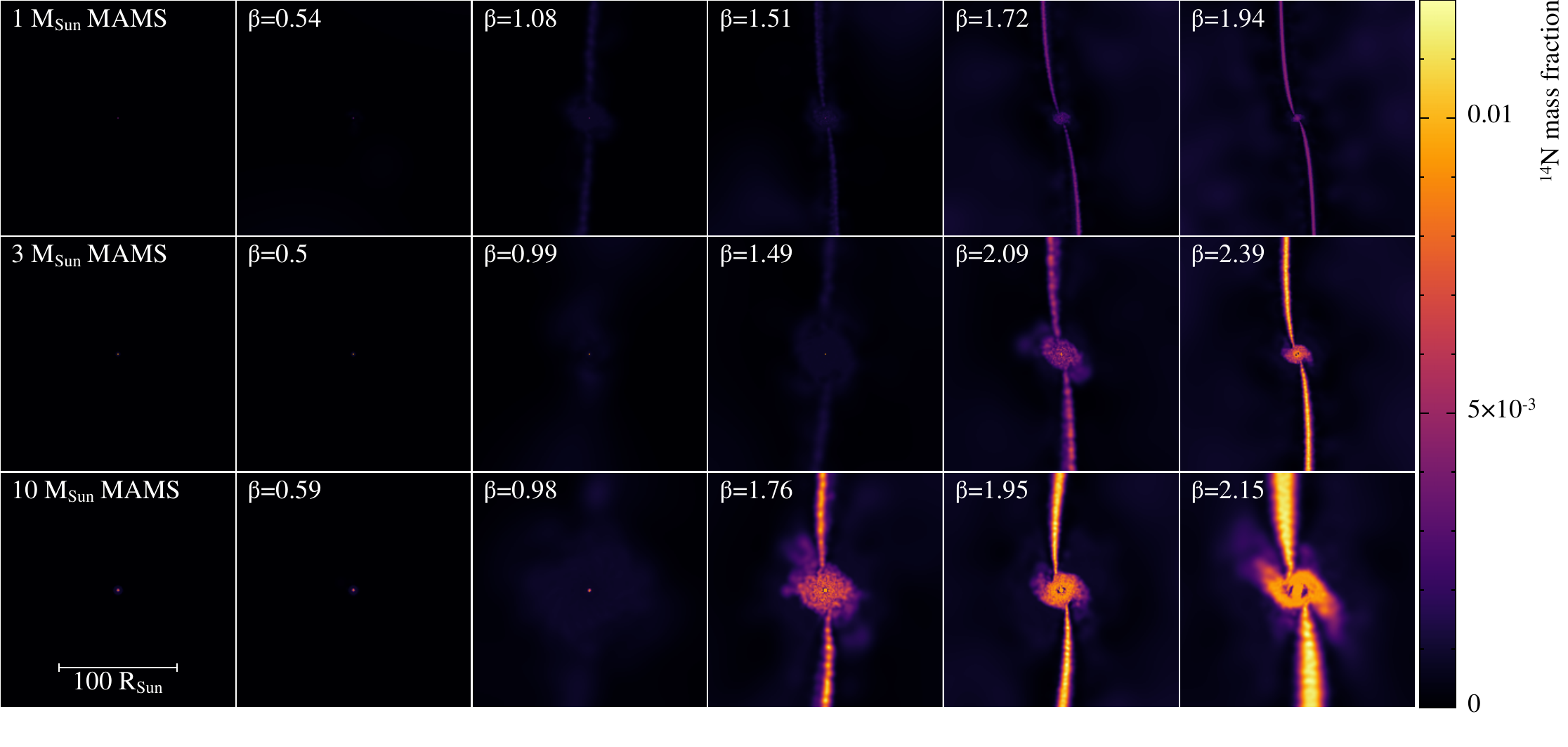}
    
   \caption{\ce{^14N} composition cross section slices in the orbital plane of $1\,\mathrm{M}_\odot$, $3\,\mathrm{M}_\odot$, and $10\,\mathrm{M}_\odot$ MAMS models at $4.8$ days, $7.3$ days, and $9$ days since disruption, respectively (\textit{top} to \textit{bottom}).  Each row shows the models at different penetration factors, with the first column showing the models before the disruption.  Stronger disruption events result in more material being removed from the centre of the star, which eventually ends up in streams.  Some material in the streams falls back onto the remnant. }
   \label{fig:compn14}
\end{figure*}

\begin{figure*}

 \includegraphics[width=\textwidth]{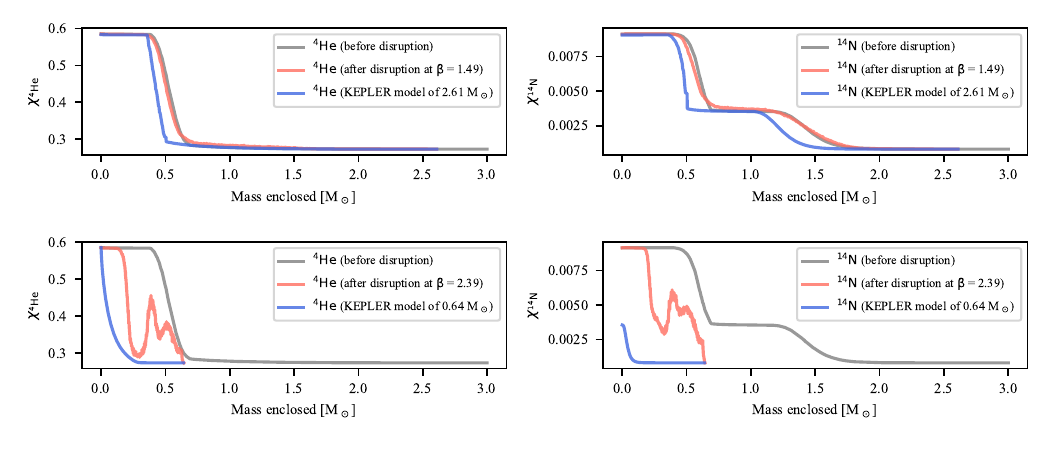}
 \caption{\ce{^4He} and \ce{^14N} mass fraction as a function of the enclosed mass of remnants formed by disrupting $3\,\mathrm{M}_\odot$ MAMS model with $\beta=1.49$ (\textit{top}) and $\beta=2.39$ (\textit{bottom}).  The green line illustrates the radially-binned mass fraction of the $3\,\mathrm{M}_\odot$ before disruption, the blue line represents the remnant formed after disruption, and the red line shows the mass fraction obtained from a \textsc{Kepler} model of the same mass.  \textsc{Kepler} models are evolved until the central \ce{^4He} mass fraction matches that of the remnant.  Notably, the remnant retaining $\sim 87\%$ of the initial model's mass experiences slight composition mixing.  The remnant retaining $\sim 21\%$ exhibits higher composition mixing. 
 Additionally, the \textsc{Kepler} models display lower mass fractions at lower enclosed masses for both these models.  \ce{^14N} content in \textsc{Kepler} model is $0.38$ times lower than that of the remnant formed with $\beta=2.39$.}

   \label{fig:comp_he_n}
\end{figure*}

Figure~\ref{fig:compn14} shows the \ce{^14N} abundance perpendicular to the orbital plane for $1\,\mathrm{M}_\odot$, $3\,\mathrm{M}_\odot$, and $10\,\mathrm{M}_\odot$ MAMS models (\textit{top} to \textit{bottom}) for a range of penetration factors (\textit{across}), at $4.8$, $7.3$, and $9$ days post-disruption. We see that as the strength of the encounter increases, there is more mixing inside the star as the disruption can affect the deeper layers of the model. We note that in the disruptions where mass loss is $\lesssim 10\%$ of the initial model's mass, there is no mixing inside the star. As the star approaches the SMBH at a closer distance, more material is mixed. Moreover, the streams have more material stripped from the central regions as the disruption encounter becomes more violent. 

Figure~\ref{fig:comp_he_n} shows the \ce{^4He} and \ce{^14N} mass-fractions as a function of the enclosed mass of remnants formed by disrupting a $3\,\mathrm{M}_\odot$ MAMS model with $\beta=1.49$ (\textit{top}) and $\beta=2.39$ (\textit{bottom}).  The green line illustrates the radially-binned mass fraction of the $3\,\mathrm{M}_\odot$ before disruption, the blue line represents the remnant formed after disruption, and the red line shows the mass fraction obtained from a \textsc{Kepler} model of the same mass.  \textsc{Kepler} models are evolved until the central \ce{^4He} mass fraction matches that of the remnant.  We see that the $\beta=1.49$, where the model lost $\sim 13\%$ mass, the mass fraction of the remnant is slightly higher than the before disruption mass-fraction for mass enclosed $>0.7\,\mathrm{M}_\odot$ due to mixing.  The \textsc{Kepler} model of the same mass has overall lower mass fractions of \ce{^4He} and \ce{^14N}. 

The remnant formed by disrupting this star, $\beta=2.39$ results in a mass loss of $\sim 83\%$. 
 There is more mixing in the remnant.  We can see \ce{^4He} enrichment in the envelope.  The \textsc{Kepler} model of the same mass has lower mass-fraction than the remnant.  \ce{^14N} content in \textsc{Kepler} model is $0.38$  times lower than that of the remnant.

\subsection{Evolution of remnants}
\label{sec:evolution_HR}

\begin{figure*}

    \includegraphics[width=0.33\textwidth]{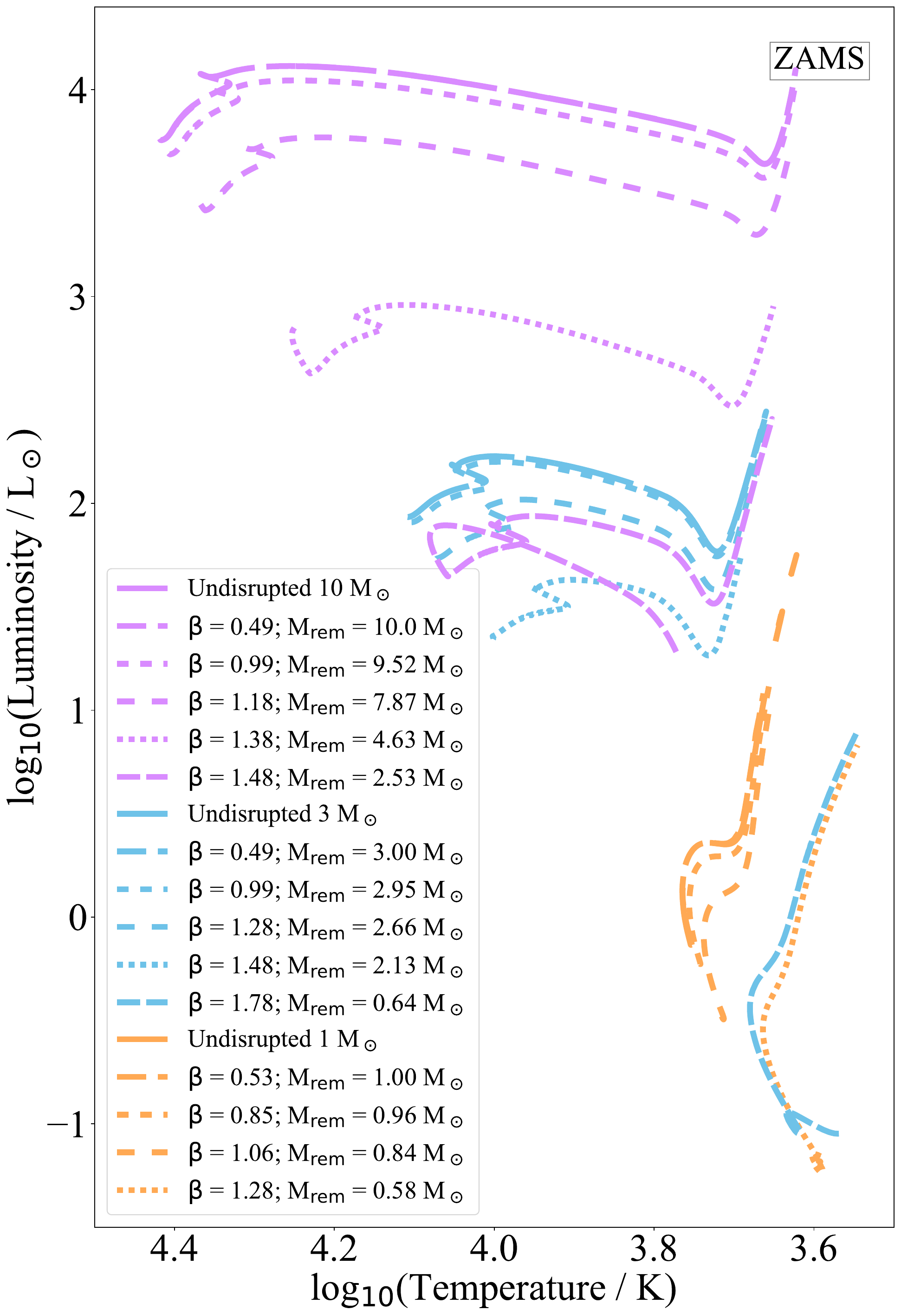}
    \includegraphics[width=0.33\textwidth]{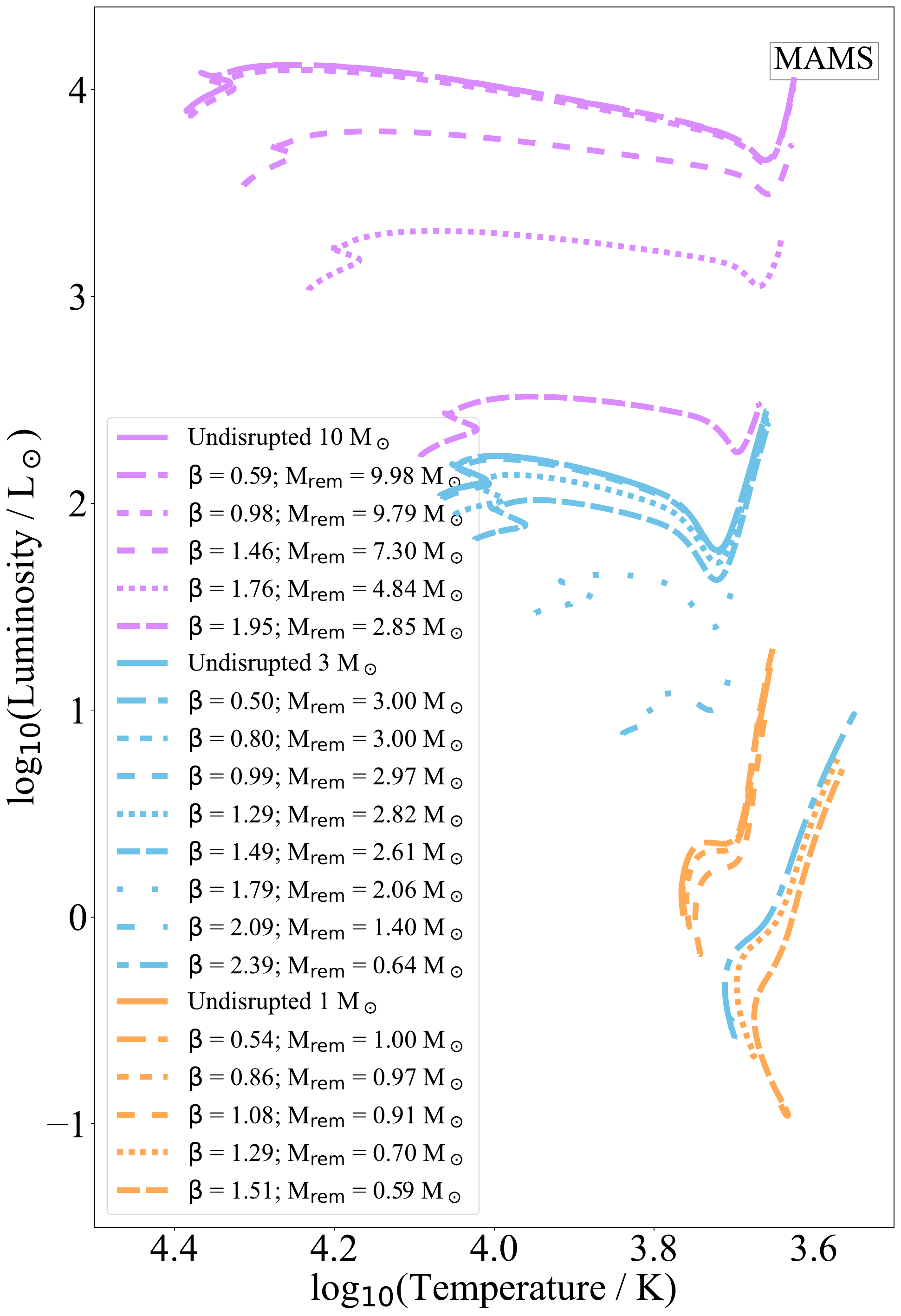}
    \includegraphics[width=0.33\textwidth]{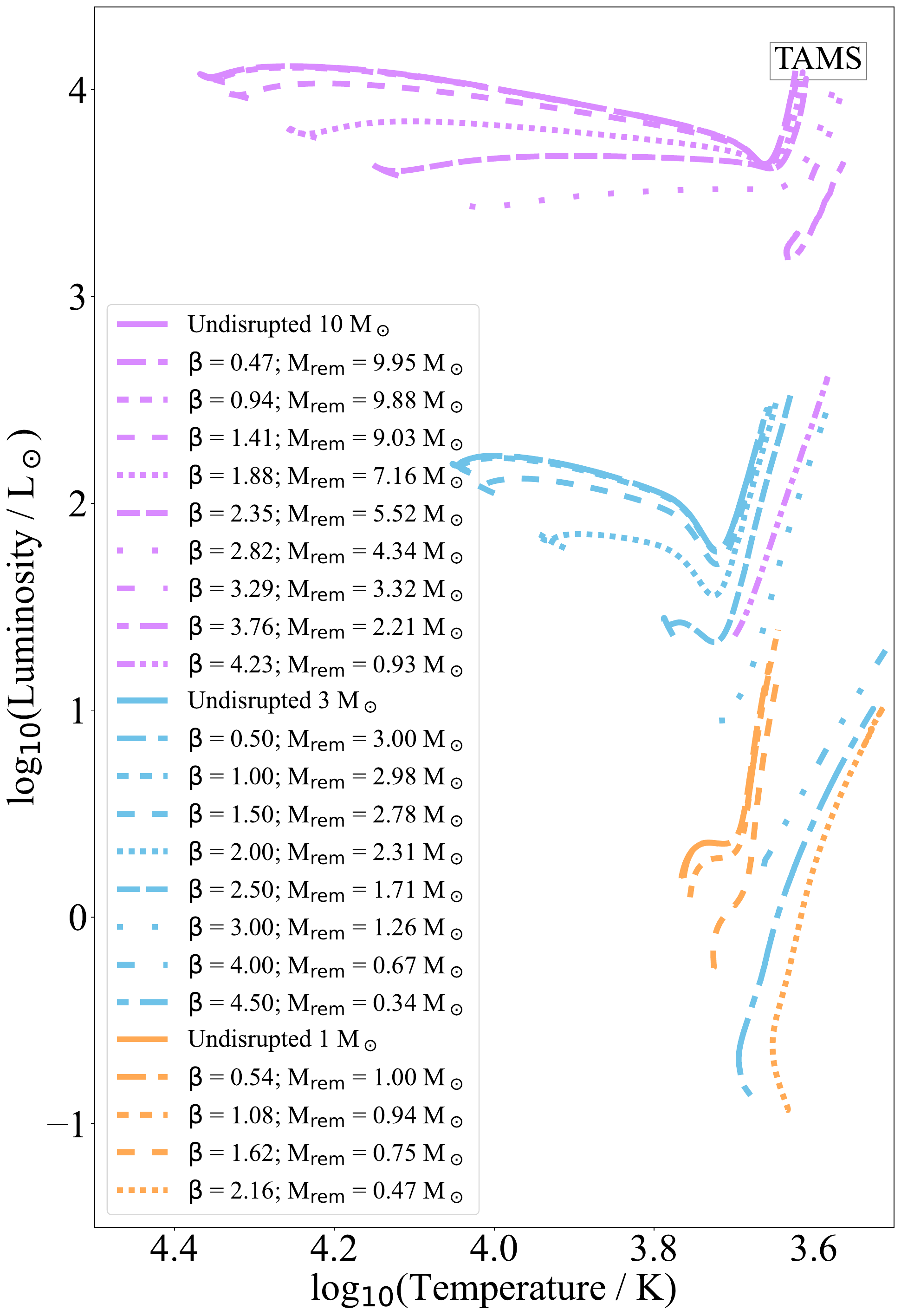}
    \caption{The \textit{leftmost}, \textit{central} and \textit{rightmost} panels show the H-R diagrams for ZAMS, MAMS and TAMS of remnants formed by disrupting $1\,\mathrm{M}_\odot$, $3\,\mathrm{M}_\odot$, and $10\,\mathrm{M}_\odot$ at a range of $\beta$ values after the Kelvin-Helmholtz time.  The evolution of remnants of similar mass formed from ZAMS models is not significantly different.  However, MAMS and TAMS remnants exhibit higher luminosity and temperature for those formed from a higher-mass star.  }
    \label{fig:HR_models}
\end{figure*}

Figure~\ref{fig:HR_models} shows the evolution of all the models following the disruption event after their respective Kelvin-Helmholtz time.  The simulations were carried out until the core ignited core helium burning. Models with mass $< 0.5\, \mathrm{M}_\odot$ posed challenges for re-evolution in \textsc{Kepler} and therefore are not shown in the plot. 

From ZAMS models (\textit{left} panel of Figure~\ref{fig:HR_models}),  we see that the remnants formed by disruption evolve on a track that is nearly identical to an undisrupted star with the same mass as the remnant.  For example, the disruption  of the $10\,\mathrm{M}_\odot$ ZAMS model with $\beta=1.48$ and the $3\,\mathrm{M}_\odot$ ZAMS model with $\beta = 1.28$ end up with masses of $2.58\,\mathrm{M}_\odot$ and $2.66\,\mathrm{M}_\odot$, respectively.  Their tracks (purple dashed line and blue dashed lines lying close to each other at $\log(Luminosity /L_\odot) \sim 2$; see legend) are nearly identical (although the slightly lower mass model exhibits correspondingly lower luminosity and effective temperature, as expected).  Similarly, the remnant of the $3\,\mathrm{M}_\odot$ ZAMS model with $\beta = 1.78$ and the remant of the $1\,\mathrm{M}_\odot$ ZAMS model with $\beta=1.28$ have masses of $0.64\,\mathrm{M}_\odot$ and $0.58\,\mathrm{M}_\odot$, respectively.  They follow similar evolution tracks (blue and orange lines lying close to each other in the bottom right of the diagram).  This is because the composition of the star is unchanged by the disruption process. 

For MAMS models (\textit{middle} panel of Figure~\ref{fig:HR_models}), the situation is different.  The remnants formed from disrupting a higher mass star have higher luminosity and effective temperature in comparison to undisrupted stars of the same mass.  For example, the remnant of the disruption of a $10\,\mathrm{M}_\odot$ model with $\beta = 1.95$ (lowest purple line in the middle panel), has higher luminosity than a $3\,\mathrm{M}_\odot$ star (solid blue line in middle panel) despite having lower mass ($2.85\,\mathrm{M}_\odot$).  Hence, in this case, we have produced an over-luminous star (see Section~\ref{sec:discussion_sec}).  Another example is disruption of a $3\,\mathrm{M}_\odot$ MAMS model with $\beta=2.39$ which results in a remannt with higher luminosity than the disruption of a $1\,\mathrm{M}_\odot$ MAMS model with $\beta=1.29$ despite having a lower mass.

For TAMS models (\textit{right-most} panel of Figure~\ref{fig:HR_models}), the situation is similar to the MAMS models.  For example, the remnant of $10\,\mathrm{M}_\odot$ TAMS model with $\beta=4.23$ (last purple line lying in the region of blue lines) has higher luminosity than a $1\,\mathrm{M}_\odot$ star which never encountered an SMBH.  This remnant is also more luminous compared to a remnant of $1.26\,\mathrm{M}_\odot$ formed by disrupting $3\,\mathrm{M}_\odot$ TAMS model with $\beta=3$ (dotted blue line after the last purple line).  The disruption of a $3\,\mathrm{M}_\odot$ TAMS model with $\beta=4.5$ yields a remnant that has higher luminosity than the $1\,\mathrm{M}_\odot$ model with $\beta=2.16$ despite the former having a smaller mass (blue and orange lines in the right corner).

\begin{figure*}
    \centering
    \includegraphics[width=\textwidth]{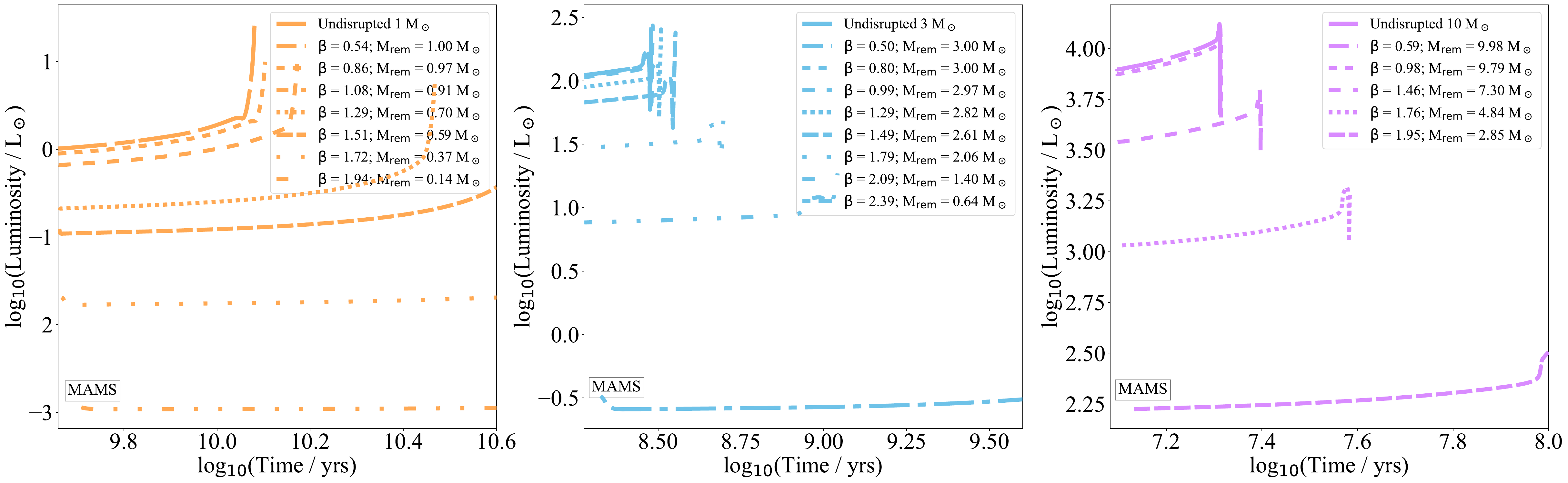}
    \caption{Luminosity as a function of time for MAMS models.  The \textit{leftmost}, \textit{central} and \textit{rightmost} panels correspond to remnants of $1\,\mathrm{M}_\odot$, $3\,\mathrm{M}_\odot$, and $10\,\mathrm{M}_\odot$ stars respectively.  The models that lose more mass tend to have longer main-sequence lifetimes.   }
    \label{fig:lumVsTime}
\end{figure*}

We also found that disrupted stars have longer lifetimes. Figure~\ref{fig:lumVsTime} shows the luminosity as a function of time for remnants of $1\,\mathrm{M}_\odot$, $3\,\mathrm{M}_\odot$, and $10\,\mathrm{M}_\odot$ MAMS models.  The remnants can live longer on the main-sequence as the mass of the remnant decreases.

\begin{figure}
    \centering    
    \includegraphics[width=\columnwidth]{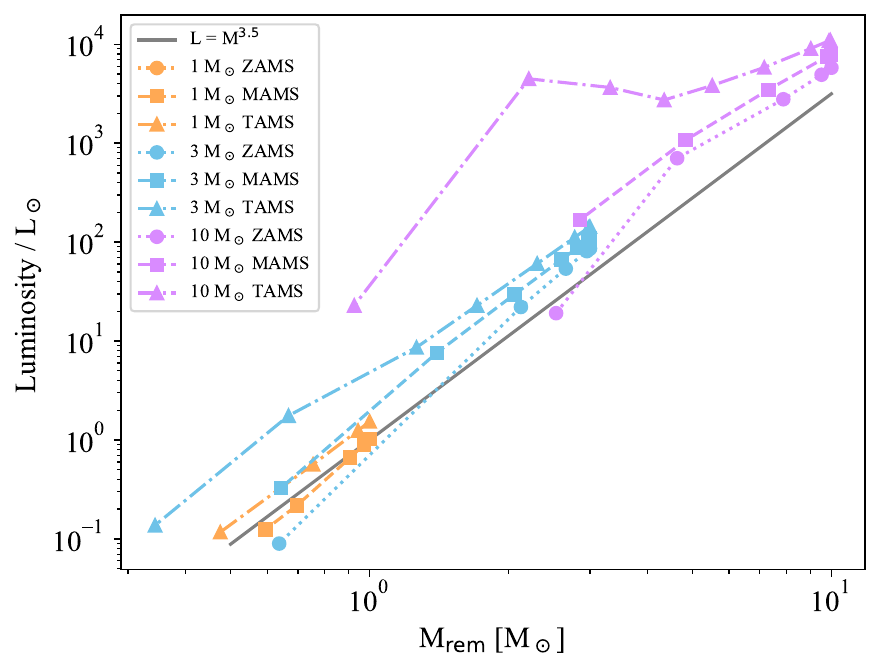}
    \caption{Luminosity at Kelvin-Helmholtz time as a function of the mass of remnant.  The grey line represents the established relationship for main-sequence stars \citep{Kuiper1938}.  Disrupted stars do not follow this relationship.}
    \label{fig:mass_vs_lum}
\end{figure}

A relationship between the luminosity of the star and its mass has been noted for main-sequence stars \citep{Kuiper1938}.  Figure~\ref{fig:mass_vs_lum} shows this relationship for our remnants at the Kelvin-Helmholtz time.  As we can see, the remnants do not follow this relationship, showing that these stars are unlike undisrupted stars.


\subsection{Nitrogen enrichment}
\begin{figure}
    \centering   
    \includegraphics[width=\columnwidth]{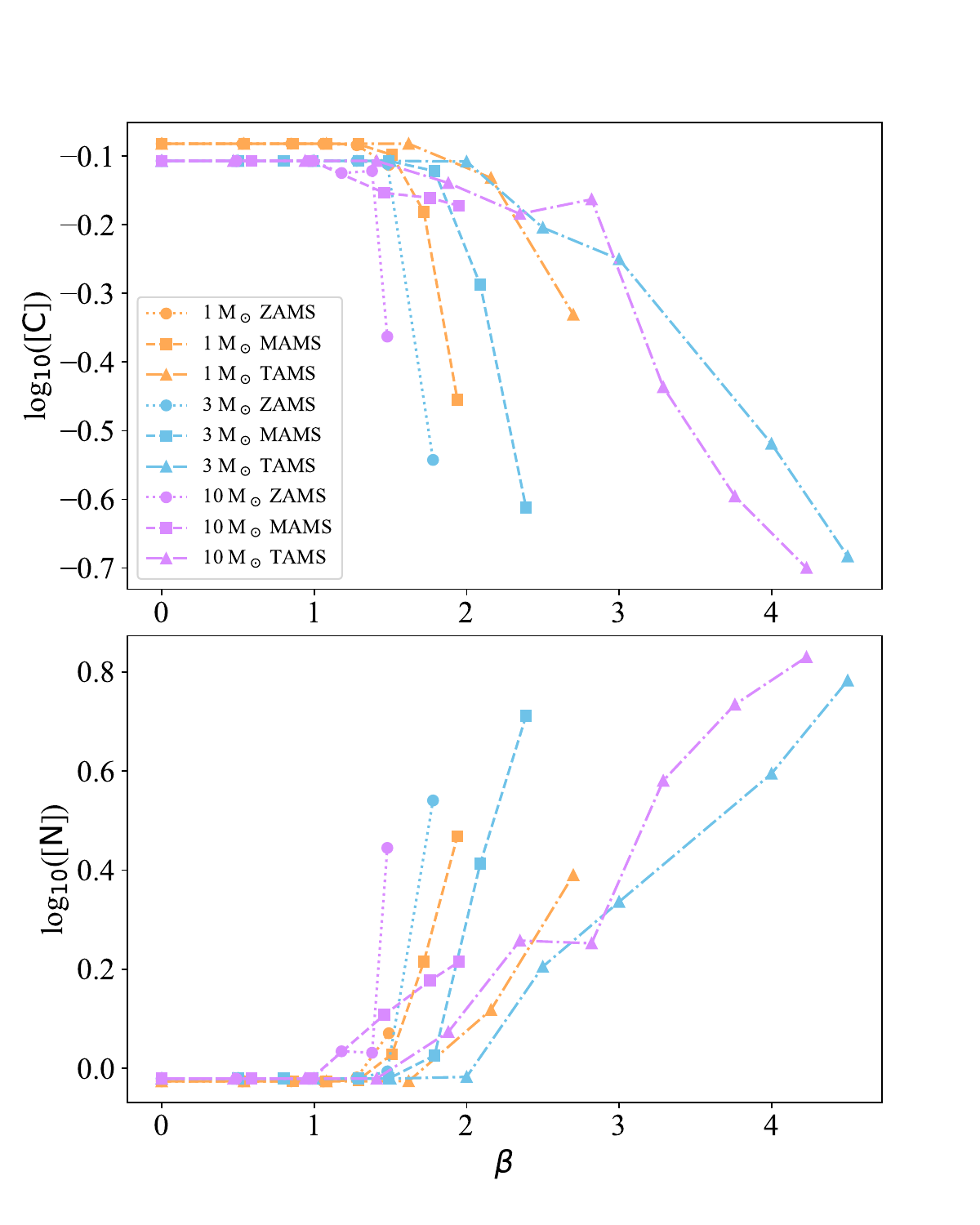}
    \caption{\ce{^12C} abundance (\textit{top}) and \ce{^14N} abundance (\textit{bottom}) at the surface of the remnant as function of $\beta$ at a Kelvin-Helmholtz time after the TDE.  
 With increasing $\beta$, \ce{^14N} is enriched, whereas \ce{^12C} is depleted, leading to significant changes in the observable [C/N] ratio.  }
    \label{fig:nh_ratio}
\end{figure}

Figure~\ref{fig:nh_ratio} shows the \ce{^12C} and \ce{^14N} abundance at the surface of the remnant as a function of penetration factor at the Kelvin-Helmholtz time.  ZAMS models have the least \ce{^14N} enrichment and some enrichment does take place for $\beta > 1$.  
 The remnants of more evolved initial stars have the most amount of enrichment.  The enrichment becomes significant for deeper encounters. This would lead to significant changes in the observable [C/N] ratio and provide a signature for detecting TDE survivor stars. 

\subsection{Comparison with non-disrupted stars of same mass}
\label{sec:Compare_models}
\begin{figure*}
    \centering    
    \includegraphics[width=\textwidth]{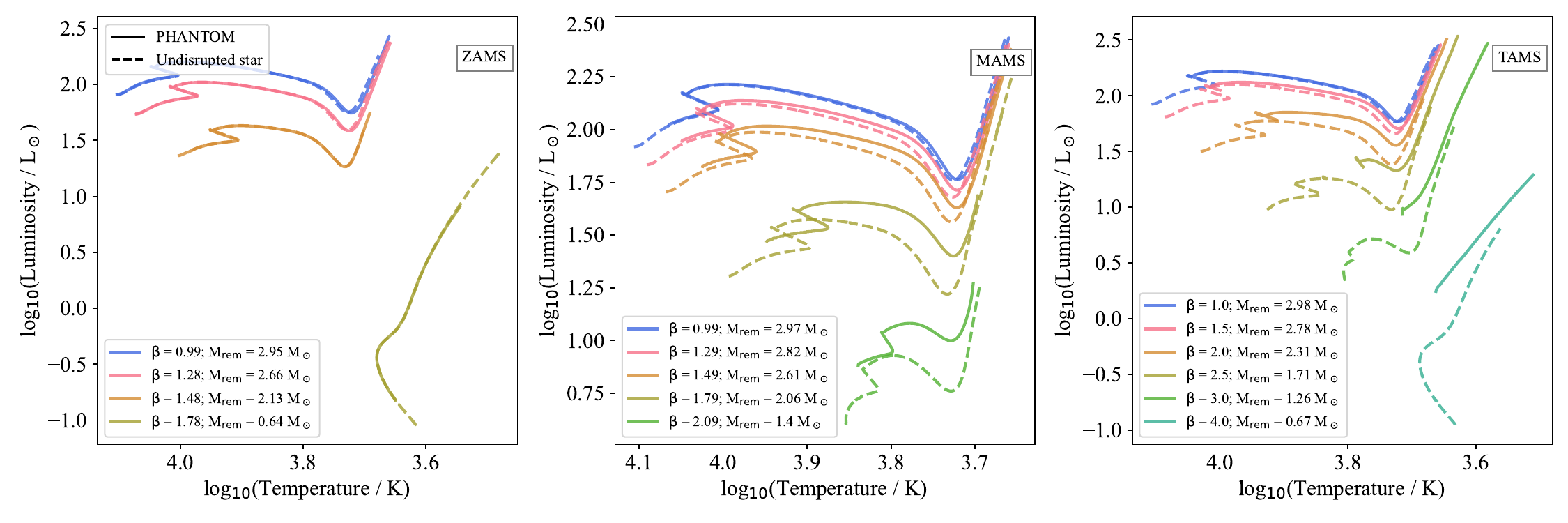}
    \caption{H-R diagram of $3\,\mathrm{M}_\odot$ remnants for ZAMS, MAMS and TAMS stages shown in \textit{leftmost}, \textit{central} and \textit{rightmost} panels respectively.  Solid lines correspond to the evolution of remnants after the Kelvin-Helmholtz time.  Dashed lines correspond to post hydrogen ignition evolution of models evolved in \textsc{Kepler} having the same mass as the remnant models.  ZAMS remnants have similar evolution to the \textsc{Kepler} models.  While MAMS and TAMS remnants have higher luminosity and effective temperature compared with \textsc{Kepler} model of the same mass.  }
    \label{fig:evolution_comparison}
\end{figure*}

Figure~\ref{fig:evolution_comparison} shows the H-R diagrams of $3\,\mathrm{M}_\odot$ remnants at ZAMS, MAMS and TAMS (\textit{across}).  Solid lines correspond to models that were disrupted in \textsc{Phantom} and then mapped back into \textsc{Kepler}.  The dashed lines correspond to the \textsc{Kepler} models of the same mass which never encountered an SMBH.  We can see that the ZAMS models follow similar evolution, but for MAMS and TAMS, the disrupted models have higher luminosity and higher effective temperature.  The difference in evolution becomes significant as we move towards lower mass remnants.

\subsection{Comparison with stars that had their outer layers removed}
We wanted to understand how different the remnants are from the original star with simply the outer layers removed.  To achieve this, we used the mapped files of our remnants but used the interpolated composition from the original undisrupted star for the remnant mass (referred to as stripped star). 

\begin{figure}
    \centering    
    \includegraphics[width=\columnwidth]{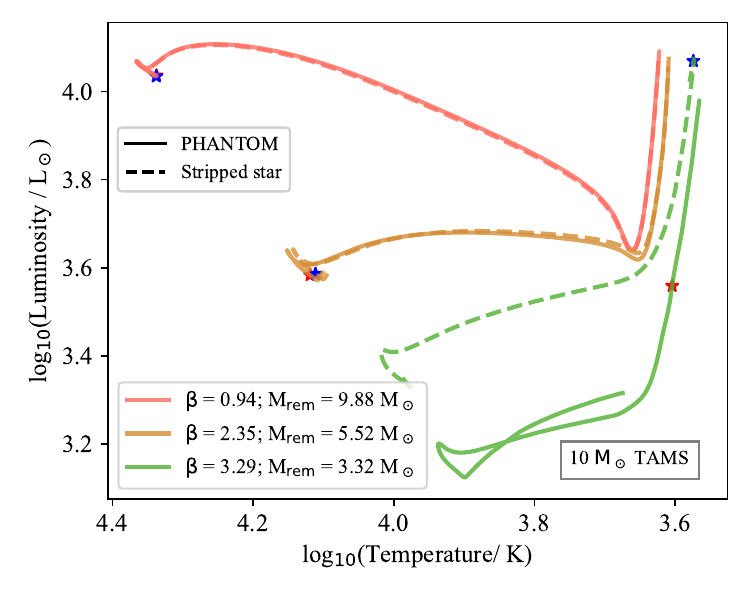}
    \caption{H-R diagram of $10\,\mathrm{M}_\odot$ TAMS remnants for a range of $\beta$ values.  We also plot the evolution of stars which would have only been stripped by the SMBH.  All models are plotted $40,000$ years since the start of \textsc{Kepler} evolution.  We show the Kelvin-Helmholtz time as star points for each model.    We see that stripped stars are more luminous compared with our remnants.  }
    \label{fig:stripped_stars}
\end{figure}

Figure~\ref{fig:stripped_stars} shows the H-R diagram of $10\,\mathrm{M}_\odot$ models.  The dashed lines represent the stripped models, while solid lines are the remnants.  We show the evolution after $40\mathord,000$ years. We also show the Kelvin-Helmholtz time of each model (starred points).  We see that the more a star is disrupted, the lower its luminosity is compared with a stripped star.  For example, $\beta=3.29$ remnant has lower luminosity than the stripped star of the same mass.  In the encounters where the star lost only $\sim 45\%$ of its mass, the evolution of the remnant is similar to the stripped star.  We found a similar behaviour for the MAMS models. 

\begin{figure}
    \centering    
    \includegraphics[width=\columnwidth]{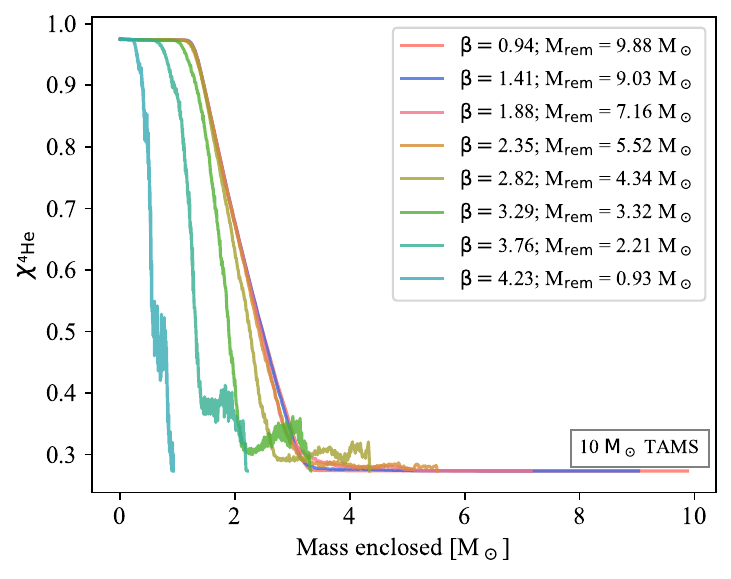}
    \caption{Mapped \ce{^4He} from \textsc{Phantom} to \textsc{Kepler} post-disruption as function of enclosed mass for $10\,\mathrm{M}_\odot$ TAMS models for a range of penetration factors.  Pre-disruption composition is retained at the centre of the remnant.  }
    \label{fig:homoge_evo}
\end{figure}

Fallback causes a H-rich surface layer, unlike just stripping the outer layers.  Yet we retain the pre-disruption composition in the core. Figure~\ref{fig:homoge_evo} shows mapped \ce{^4He} from \textsc{Phantom} to \textsc{Kepler} post-disruption as a function of enclosed mass for $10\,\mathrm{M}_\odot$ TAMS models for a range of penetration factors.  This shows that disrupted stars are unlike stripped stars for more deeply penetrating encounters.

\section{Discussion}
\label{sec:discussion_sec}
Our study has several inherent limitations. Firstly, we did not incorporate rotation into the mapping within \textsc{Kepler}. Additionally,  we did not disrupt stars present in the Nuclear Star Cluster of the Milky Way. Furthermore, the boundary distinguishing partial and full TDEs requires further exploration, necessitating an in-depth understanding of how the spin of the star and SMBH influence such encounters.

\subsection{Comparison to other work}

\subsubsection{Properties of remnants}

\citet{Coughlin2022nov} argued that the the tidal stripping of the star would take place for $\beta \gtrsim 0.6$ which is in agreement with our numerical results.

Partial tidal disruption events manifest as violent occurrences, inducing oscillations in the remnant star close to the fundamental mode, accompanied by an increase in central density near the pericentre (Figure~\ref{fig:max_den_time_oscill}). \citet{Guillochon2013} also identified post-pericentre oscillations in their tidal disruption simulations of polytropic models. The small increase in density before pericentre passage is not obvious in their plots since they use a log scale. 

\citet{NixonC2022}, utilising polytropic models, found that the central density of their star increased near the pericentre.  They argued that the first spike is a result of vertical oscillation, while the second spike is caused by in-plane caustics.  
 
TDEs lead to the stripping of stars, resulting in mass moss, as has been reported by several authors \citep{Tal2005,Lodato2009,Guillochon2013,Manukian2013,Goicovic2019,Ryu12020,Ryu2020b,Ryu32020}.  The amount of mass lost increases with the impact parameter ($\beta$) as shown in Figure~\ref{fig:remnant_mass_vs_beta}.  Sometimes a star might be completely destroyed but material can fall back onto itself to form a remnant as has also been found by \citep{Guillochon2013,NixonC2022}.   
Evolved stars, such as TAMS can survive disruption closer to the SMBH compared to ZAMS stars, consistent with the findings of \citet{Lawsmith2020}.  Remnants exhibit lower densities, temperatures and larger radii after the encounter (Figures~\ref{fig:rendered_models_new} and \ref{fig:Binned_density}).  In contrast to the prediction by \citet{TalandMario2001} that the radius of the remnants would be smaller, our results suggest otherwise.  Additionally, they had predicted that disrupted stars would experience chemical mixing, which we also observed in our models (Figures~\ref{fig:compn14} and \ref{fig:comp_he_n}).
 
We observed the formation of vortices that can redistribute angular momentum and result in the central regions rotating faster (Figure~\ref{fig:rotaion_mixing}).  This observation aligns with \citet{Goicovic2019}, who disrupted a $1\,\mathrm{M}_\odot$ ZAMS \textsc{MESA} model \citep{Paxton2011}, although their vortices persisted for the duration of their simulations ($44$ hours).
 
Our remnants exhibit rigid body rotation in the central regions (close to the radius of the initial model), and differential rotation in outer layers (Figure~\ref{fig:omega_cylindrical}).  This is unlike \citet{Goicovic2019} who found that their core did not exhibit rigid rotation.  The discrepancy might be attributed to their models having an initial rigid rotation profile, whereas we assumed initially non-rotating stars.  
 
\citet{Sacchi2019} disrupted a $1\,\mathrm{M}_\odot$, $\gamma=\nicefrac53$ polytrope with $\beta=1$. Unlike our approach, they provided the initial model with initial rigid body rotation.  In their simulations, they found that a disk can form around 3 days after disruption,  tightly bound to the core and exhibiting Keplerian rotation.  This is similar to our observation, although we do not see the formation of a disk in our models.  We speculate that the discrepancy might be due to their model having an initial rotation profile close to differential rotation. Further work is required to resolve this difference.  
 
\citet{Nixon2021} also found a disk in for $\beta = 1.6$, $1\,\mathrm{M}_\odot$ ZAMS model, although they did not specify if it was bound to the star. The cause of the discrepancy with our results remains uncertain. 
They argued that the disk formed around the core is spun off from it.  In our simulations, the differentially rotating region belongs to the outer layers of the initial star that were stripped away. As the simulation progresses, material from the outer layers of the rigidly rotating core can become part of these differentially rotating layers; moreover, the streams can fall back onto the star.

\citet{Ryu32020} also calculated the angular velocity, but unlike our models, their models rotate faster in the outer layers.  They found that for the same fractional mass loss, the models that have higher initial mass get closest to break-up velocity.  We do not observe such a trend. For example, $\beta=1.29$ for a $1\,\mathrm{M}_\odot$ MAMS, and $\beta=2.39$ for a $3\,\mathrm{M}_\odot$ MAMS, form similar mass remnants, but the remnant formed from the latter model is not closer to break-up velocity than the former models' remnant.

By evolving some of our models to long times, we found that the size of the remnant increases (Figure~\ref{fig:disks1} and Figure~\ref{fig:angular_vel_wtr_time}).  Moreover, the angular momentum is redistributed with time, with differentially rotating layers rotating faster while the core rotates slower. 

Further investigation is necessary to determine the cause of discrepancies in the rotation profiles within different simulation codes.

\subsubsection{Critical $\beta$ for total disruption}

We calculated the critical $\beta$ ($\beta_\mathrm{c}$) values as $1.56$, $2.04$, $3.40$ for $1\,\mathrm{M}_\odot$ ZAMS, MAMS and TAMS models as described in Section~\ref{sec:critical_beta_sec}.  For $3\,\mathrm{M}_\odot$ model, $\beta_\mathrm{c}$ was determined as $1.91$, $2.59$, and $4.8$ for ZAMS, MAMS and TAMS stages.  In the case of $10\, \mathrm{M}_\odot$, we estimated it as $1.59$, $2.17$, and $4.49$ ZAMS, MAMS and TAMS stages, respectively.  \citet{Guillochon2013} were the first to explore $\beta_\mathrm{c}$ for polytropes with $\gamma$ values of $\nicefrac{4}{3}$ and $\nicefrac{5}{3}$ as $1.85$ and $0.9$, respectively.  As polytropic models have lower central densities, the star would not be able to survive encounters closer to the SMBH, compared with a star of higher central density. 

\citet{Lawsmith2020} utilised \textsc{MESA} stellar models and disrupted them in Newtonian physics.  For their $1\,\mathrm{M}_\odot$ ZAMS model, they determined a $\beta_\mathrm{c}$ of $1.8\pm0.1$. Their model had a radius of $0.9\,\mathrm{R}_\odot$ and $\rho_\mathrm{c}/\bar{\rho}$ of $42$, which is similar to our model.  Hence, the difference in $\beta_\mathrm{c}$ might stem from using general relativistic effects in our simulations. Their TAMS model was obtained at $8.4\,\mathrm{Gyr}$ and had a radius about $5\%$ larger and $\rho_\mathrm{c}/\bar{\rho}$ about $75\%$ higher than our model.  They determined a $\beta_\mathrm{c}$ of $7$ for this model, about twice of what we found.  This can be attributed to the difference in the radius and density of the stellar models.

Their $3\,\mathrm{M}_\odot$ ZAMS and TAMS models had radius of $1.89\,\mathrm{R}_\odot$ and $3.32\,\mathrm{R}_\odot$ with $\rho_\mathrm{c}/\bar{\rho}$ of $73$ and $1198$.  Their ZAMS $\beta_\mathrm{c}$ was about $6\%$ higher than ours. This could be either a result of a difference in $\rho_\mathrm{c}/\bar{\rho}$ or general relativistic effects.  $\rho_\mathrm{c}/\bar{\rho}$ for our TAMS model was $24\%$ lower. This could be a possible explanation for the difference in $\beta_\mathrm{c}$. 

\citet{Goicovic2019} used $1\,\mathrm{M}_\odot$ ZAMS model from \textsc{MESA}, and determined that the full disruption takes place after $\beta\geq2$.  Hence, there is a slight difference in the literature results. 

\citet{Ryu32020} used real stellar models that had spent about half of their lives on main-sequence and disrupted them in a fully general relativistic framework.  There is a small discrepancy between their $\beta_\mathrm{c}$ values and ours for the $1\,\mathrm{M}_\odot$ and $10\,\mathrm{M}_\odot$ MAMS models.  This could be due to us using hyperbolic orbits for the simulations as for deep encounters, parabolic orbits lose more mass than the hyperbolic cases as discussed in Section~\ref{app:zero_e_orbits}.  As \citet{Ryu32020} took their models from \textsc{MESA} while we utilised \textsc{Kepler}, this could also be a reason for the difference.  The small discrepancies could also be a result of different simulation methods \citep{Mainetti2017}.

We also explored the correlation between $\beta_\mathrm{c}$ and $(\rho_\mathrm{c}/\bar{\rho})$. The results are given in Eq.~\ref{eq:fit_form}. 
 \citet{Lawsmith2020} performed the fit based on the distinction in density ratio, proposing a function of the form $\beta_\mathrm{c} = 0.5\times(\rho_\mathrm{c}/\bar{\rho})^{\nicefrac13}$ for $(\rho_\mathrm{c}/\bar{\rho}) \lesssim 500$ and $\beta_\mathrm{c} = 0.39\times(\rho_\mathrm{c}/\bar{\rho})^{\nicefrac1{2.3}}$ for $(\rho_\mathrm{c}/\bar{\rho}) \gtrsim 500$.  Our $1\,\mathrm{M}_\odot$ TAMS and $10\,\mathrm{M}_\odot$ TAMS models, however, have $\rho_\mathrm{c}/\bar{\rho}$ of $569.76$ and $563.72$, respectively.  As models below a mass of $1.3\,\mathrm{M}_\odot$ have convective cores while those above this mass have radiative cores, fitting the same function to $\beta_\mathrm{c}$ would overlook the evolutionary structure of these stars. 

\citet{Lawsmith2020} used Newtonian physics in their models and observed remnant formation at higher $\beta$ values than the $\beta_\mathrm{c}$ that we determined.  Additionally, the $\rho_\mathrm{c}/\bar{\rho}$ values of their models differ from ours, owing to slight differences in stellar evolution codes.  

\citet{Ryu2020b} derived a semi-analytical formula of the form $\beta_\mathrm{c}  = 2.14\times(\rho_\mathrm{c}/\bar{\rho})^{\nicefrac13}$.  \citet{Coughlin2022nov} determined a  a functional form of $\beta_\mathrm{c}  = 0.62\times(\rho_c/\bar{\rho})^{\nicefrac13}$ and found good agreement with previous numerical simulations.  But \citet{NixonC2022} had found that the star can survive the disruption encounter for $\beta > \beta_\mathrm{c}$ for their $\gamma =\nicefrac{5}{3}$ polytrope, $\beta = 16$, while previous studies had determined a value of $\sim$ $0.9$ \citep{Guillochon2013,Mainetti2017}.  The cause of this discrepancy is unclear.

\subsubsection{Evolution of remnants }
Hydrogen burning in the low-mass stars is dominated by the PP chain, however, for more massive stars the CNO cycle starts to dominate hydrogen burning.  At first, \ce{^12C} is converted to \ce{^14N} very swiftly at the beginning of burning.  This is the CN cycle.  There is also a branch that converts \ce{^16O} to \ce{^14N} to participate in the CN cycle, however, this conversion by the ON branch requires higher temperatures.  It is usually only encountered toward the end of the evolution of a 1 solar mass star and still proceeds much slower than the initial conversion of \ce{^12C} even in higher-mass stars.  Usually, not much of that occurs at the early evolution stage we refer to as ZAMS but will be present increasingly more with stellar age and mass, so, e.g., more present in the MAMS $10\,\mathrm{M}_\odot$ star than in the MAMS $3\,\mathrm{M}_\odot$ star.  Hydrogen burning by the CNO cycle --- dominated by the CN branch --- hence is enhanced when \ce{^16O} has been converted to \ce{^14N}, which would not occur in the early evolution of low-mass stars.  Hence, if one made a low-mass star from the core of the high-mass star, the resulting larger \ce{^14N} can affect the evolution as a larger CN isotope abundance can burn hydrogen faster, i.e., a lower core temperature is needed to produce the same luminosity.  

The other key factor is the enrichment of helium.  For chemically homogeneous stars, the luminosity is proportional to the fourth power of the mean molecular weight, $\mu$ \citep{Kippenhahn1990}, i.e., $\mu^4$.  The change in mean molecular weight during hydrogen burning is dominated by the change in helium mass fraction --- the change in the CNO isotopes is negligible in comparison.  Hence the helium-rich core of a stripped evolved star is more luminous than a non-stripped star of the same mass and same core helium content, as it will have more helium in the outer layers and hence a higher average mean molecular weight.  The higher helium abundance in the outer layers also leads to a lower electron scattering opacity, however, our H-R diagram shows that the stars do become redder, e.g., well-seen for the $10\,\mathrm{M}_\odot$ models.

 \citealt{TalandMario2001,Tal2005} had predicted that a partial tidal disruption would lead to remnants of main-sequence stars exhibiting higher luminosities for a given mass due to higher mean molecular weight and a larger convective region.  Our findings support this proposition, as evidenced by a comparison of our remnants with stars of the same mass, revealing higher luminosities and temperatures (Figure~\ref{fig:evolution_comparison}). However the remnants are less luminous compared to stars where only the outer layers are removed. This is because the remnants retain the pre-disruption composition in the core and have a H-rich surface layer (Figure~\ref{fig:stripped_stars} and Figure~\ref{fig:homoge_evo}).

Figure~\ref{fig:HR_models} shows that as the remnant mass decreases, it traverses the H-R diagram, towards lower effective temperature and luminosity.  This movement is not smooth for TAMS models. Moreover, the observed correlation between mass loss and extended main-sequence lifetimes suggests that stars experiencing greater mass loss can reside on the main-sequence for a longer duration --- a true `elixir of life' (Figure~\ref{fig:lumVsTime}).

We also found that disruption can result in depletion of \ce{^12C} and enrichment of \ce{^14N} at Kelvin-Helmholtz time as was predicted by \citet{TalandMario2001} (Figure~\ref{fig:nh_ratio}). 

\subsubsection{Bound or unbound?}

Stars with velocities greater than the local escape velocity of the galaxy can be either classified as hyper-velocity stars (HVSs) or hyper-runnaway stars (HRSs), where the former have a galactic centre origin \citep{Hills1988,Brown2015,Marchetti2018}.  An ejection rate of $10^{-2}$--$10^{-4}\,\mathrm{yr}^{-1}$ has been estimated for HVSs \citep{Brown2015,Evans2022}.  Detection of HVSs can help us to understand the galactic centre environment.

Since the first detection by \citet{Brown2005}, $20$ HVSs have been confirmed with velocities of $300$--$1\mathord,700\,\mathrm{km}\,\mathrm{s}^{-1}$ \citep{Brown2014,Brown2015,Koposov2020}, and more than $500$ candidates have been identified \citep{Boubert2018,Luna2024}.  Most of these stars are B-type stars with masses ranging from $2.5$--$4\,\mathrm{M}_\odot$ \citep{Luna2024}.  Only \citet{Koposov2020}'s A-type HVS with a velocity of $\sim 1\mathord,700\,\mathrm{km}\,\mathrm{s}^{-1}$ has an orbit that points towards a galactic centre origin and a result of Hill's mechanism \citep{Hills1988}.  The current astrometry is not precise enough to trace other HVSs to a galactic centre origin \citep{Evans2021}. 

Our remnants formed by disrupting stars on hyperbolic orbits result in velocities at infinity of $200$--$2\mathord,000\,\mathrm{km}\,\mathrm{s}^{-1}$.  This is different from the result of \citet{Ryu32020} who had found that all their remnants formed from disrupting $1\,\mathrm{M_\odot}$, $3\,\mathrm{M_\odot}$ and $10\,\mathrm{M_\odot}$ MAMS stars were bound to the SMBH on highly eccentric orbits.  They used parabolic orbits while we used hyperbolic orbits for our paper.  If we compare our results from the zero energy orbits discussed in Section~\ref{app:zero_e_orbits}, we note that only $\beta=1.49$ disruption results in an unbound remnant with a velocity $\sim600\,\mathrm{km}\,\mathrm{s}^{-1}$.  All the subsequently bound models follow highly eccentric orbits around the SMBH post-disruption.  Some of our remnants formed from disrupting $10\,\mathrm{M}_\odot$ models were also bound to the SMBH on highly eccentric orbits. These could have repeated disruptions with the SMBH \citep{Antonini2011, Ryu32020}. \citealt{Manukian2013,Gafton2015} had found that single stars on parabolic orbits before partial disruption could have velocities of a few hundred $\mathrm{km}\,\mathrm{s}^{-1}$ after interaction with a SMBH, due to asymmetry in the mass loss.  This can explain our remnant formed by disrupting a star on a zero-energy orbit with $\beta=1.49$. 

Disruptions on hyperbolic orbits could result in HVSs. 
 Though \citet{Ryu32020} had argued that unbound remnants can experience a change in their angular momentum from gravitational encounters which can alter their trajectories.  With a logarithmic potential, post sphere of influence, the unbound remnants could reach a turning point, and end up in the galactic centre but with a small possibility of another disruption with the SMBH. 

\subsection{Possible observational candidates}

Due to difficulty in observing stars in the galactic centre, there are no present partial TDE star candidates. 
 Here, we propose a few stellar objects that might be a result of an encounter with a SMBH.

About six G-objects have been detected within $0.04$ pc of our galactic centre \citep{Gillessen2012,Phifer2013,Witzel2017,Ciurlo2020}. 
 \citet{Ciurlo2020} argued that $\mathrm{Br}\gamma$ is the main identified feature of these objects which results from ionisation of dust and is independent of the mass of the possible central object which could be a star. There is a lack of $L'$-band emissions but it does not rule out the possibility of a remnant embedded within these ionised envelopes.  A few possible explanations have been proposed for explaining the formation of these objects, such as gas or dust clouds \citep{Pfuhl2015}, a star with stellar wind \citep{Scoville2013}, or a star having an evaporating protoplanetary disk around a young star \citep{Murray-Clay2012}, or a product of stellar binary collision \citep{Prodan2015}.  As the dynamics are similar to a stellar object, we propose that partial tidal disruption of a star could have resulted in such objects.  As our simulations evolve, we see that the star increases in size (object identified as being a star based on Section~\ref{sec:method_rem}), forming a bound envelope around the dense central region.  As we looked at some of our models only about $100$ days, the exact comparison between the sizes of G-objects and our remnants is not possible.  We also did not consider cooling in our models which could have an impact on the structure of these outer layers. Furthermore, our stars had no rotation in the beginning.  All these possibilities need to be further explored.

\citet{TalandMario2001} had identified IRS~7, an M supergiant in the galactic centre as a candidate for a partial TDE.  It was found to have abundances that were consistent with the dredge-up of CNO cycle products but required deep mixing in excess compared to standard models of red giant stars \citep{Carr2000}.  But recently, \citet{Guer2022} showed that the chemical abundances are similar to rotating stellar models.  We found tidal disruption can cause mixing in the star. Moreover, the remnants can rotate faster compared to the original star (see Section~\ref{sec:rotational_properties}).  It would be interesting to compare the effects of rotation \citep{Heger2000,Maynet2000} induced by disruption on the evolution of a disrupted star, and see if they can explain IRS~7 or not. 

\citet{Schiavon2017} found N-enriched stars in the inner region of the Milky Way galaxy.  Some of these stars might be products of tidal disruption from the SMBH.  As shown in Figure~\ref{fig:nh_ratio}, remnants of partial TDEs can have Nitrogen enrichment with Carbon depletion at Kelvin-Helmholtz time after the disruption.  But the detected N-enriched stars are uniformly spread out and move in a way that cannot be distinguished from other stars within the same volume  \citep{Schiavon2017}, kinematical analysis is required to understand if this can be a formation mechanism for such stars. 

\section{Conclusion}
\label{sec:conclusion_sec}
In this paper, we obtained $1\,\mathrm{M}_\odot$, $3\,\mathrm{M}_\odot$, and $10\,\mathrm{M}_\odot$ models at zero-age, middle-age and terminal-age main-sequence for a range of $\beta$ values using the 1D stellar evolution code \textsc{Kepler}.  Then we disrupted these models in 3D general relativistic hydrodynamics using \textsc{Phantom}, and finally mapped these back into \textsc{Kepler} to understand the post Kelvin-Helmholtz evolution. 

Our key results are 
\begin{enumerate}
    \item $1\,\mathrm{M}_\odot$, $3\,\mathrm{M}_\odot$, and $10\,\mathrm{M}_\odot$ stars can all survive deep encounters ($\beta \gtrsim 1$), and result in formation of remnants. 
     \item These remnants rotate rigidly in the core and exhibit differential rotation outside. This differentially rotating region forms a large circumstellar envelope. 
    \item Around pericentre, there is an increase in central density, but as stripped material moves away from the star, the density decreases. In the end, this leaves behind remnants with lower densities and temperatures. 
    \item Remnants can live on the main-sequence for much longer than the original star they were disrupted from. Thus partial disruption of a star supplies the ``elixir of life''! 
    \item Remnants formed from different mass ZAMS stars evolve similarly to undisrupted stars of a lower mass. But for MAMS and TAMS, the remnants of higher mass stars have higher luminosities and temperatures than an undisrupted star of the same mass. 
    \item We find that partially disrupted stars can show surface enrichment of \ce{^14N}, with more enrichment occurring for deeper encounters. 
    \item Remnants are more luminous compared with stars of the same mass that were never disrupted by a SMBH. 
    \item Remnants are less luminous compared with stars where only the outer layers were removed. This is because the remnants retain a H-rich outer layer.
    
\end{enumerate}

\section*{Acknowledgements}
This work was supported by resources awarded under Astronomy Australia Ltd's ASTAC merit allocation scheme on the OzSTAR national facility at Swinburne University of Technology and gadi at the National Computing Initiative (NCI). OzSTAR receives funding from the Australian Government and the Victorian Government. This work was also supported by the Australasian Leadership Computing Grants scheme, with computational resources provided by NCI Australia, an NCRIS enabled capability supported by the Australian Government. We acknowledge awards of developer time from the Australian Astronomy Data and Computing Service (ADACS). We used {\sc splash} \citep{Pricesplash2007} for plots and visualisations from the SPH code. We also utilised {\sc numpy} \citep{Numpy2020}, {\sc matplotlib} \citep{Hunter2007}, and {\sc scipy} \citep{Scipy2020} for analysis and plotting. 

%%%%%%%%%%%%%%%%%%%%%%%%%%%%%%%%%%%%%%%%%%%%%%%%%%
\section*{Data Availability}
Our snapshots, parameter files, and evolution scripts have been uploaded to zenodo, and will be published on paper acceptance.

%%%%%%%%%%%%%%%%%%%% REFERENCES %%%%%%%%%%%%%%%%%%

% The best way to enter references is to use BibTeX:

\bibliographystyle{mnras}
\bibliography{tidal} %  bibtex file is called tidal.bib

% Alternatively you could enter them by hand, like this:
% This method is tedious and prone to error if you have lots of references
%\begin{thebibliography}{99}
%\bibitem[\protect\citeauthoryear{Author}{2012}]{Author2012}
%Author A.~N., 2013, Journal of Improbable Astronomy, 1, 1
%\bibitem[\protect\citeauthoryear{Others}{2013}]{Others2013}
%Others S., 2012, Journal of Interesting Stuff, 17, 198
%\end{thebibliography}

%%%%%%%%%%%%%%%%%%%%%%%%%%%%%%%%%%%%%%%%%%%%%%%%%%

%%%%%%%%%%%%%%%%% APPENDICES %%%%%%%%%%%%%%%%%%%%%

\appendix

\section{Determining what belongs to the star}
\label{app:sectionA}
We found that defining a remnant only by the bound material included parts of the tidal streams. In order to separate material in the streams from material in the remnant itself, we noticed a distinct temperature difference between material in the (colder) stream compared to material in the remnant, with a separation at around $8\mathord,000\,\mathrm{K}$.

\begin{figure}
    \centering
    \includegraphics[width=\columnwidth]{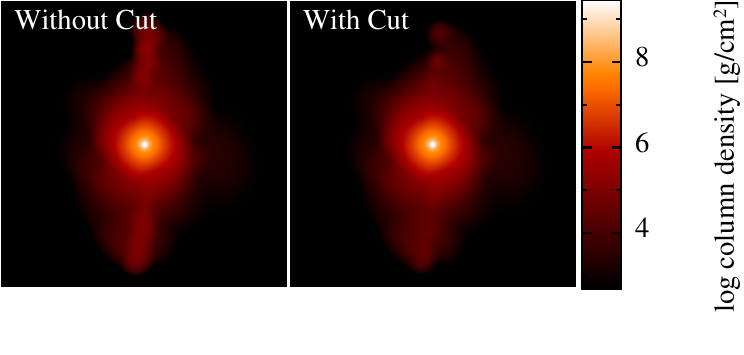}
    \caption{The before (\textit{left}) and after (\textit{right}) the cut based on temperature was implemented for a $3\,\mathrm{M}_\odot$ MAMS disrupted with $\beta=1.5$ at $8$ days since the beginning of the simulation.  We can see that the streams are correctly identified and ignored from the model. }
    \label{fig:temperature_cut}
\end{figure}

To remove the streams, we binned the bound stellar material as a function of temperature and calculated the mass in each bin.  We then calculated the mean ($\mu$) and standard deviation ($\sigma$) in the mass of these bins.  We determined the bin which had a count greater than the mean by $3\,\sigma$.  We then refine this temperature cut by finding the last bin where the mass of the bin is below a certain percentage ($1\%$ for our analysis) of the previously identified cut based on $\mu$ and $\sigma$.  This ensured that anything that was part of the stream was ignored in our analysis. Figure~\ref{fig:temperature_cut} shows the temperature cut that was performed for $3\,\mathrm{M}_\odot$ MAMS, $\beta=1.5$ at $8$ days.  The \textit{left} panel shows the material that was determined as being part of the remnant.  The \textit{right} panel shows the model after the temperature cut. This method removes most of the stream around the remnant. 
\begin{figure}
    \centering
    \includegraphics[width=\columnwidth]{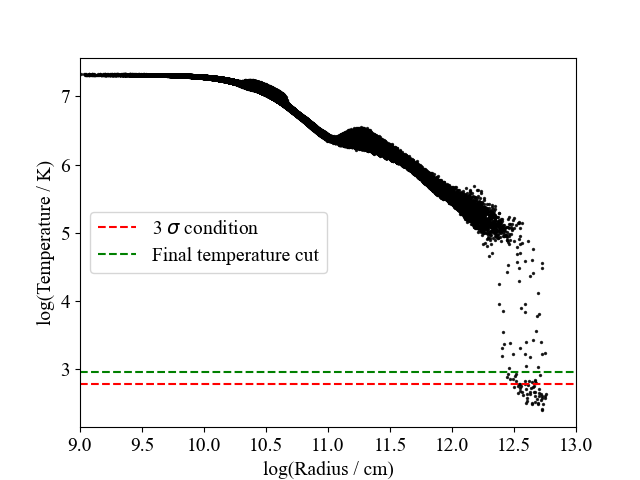}
    \caption{Temperature as function of radius of $3\,\mathrm{M}_\odot$ MAMS disrupted with $\beta=1.5$ at $8$ days.  We performed a cut in temperature on particles that were found to be bound to the remnant.  The red line shows the temperature determined based the $3\,\sigma$ condition. 
 Then we refine this cut resulting in the final cut used for analysis. }
    \label{fig:tempcutt}
\end{figure}

\section{Re-collapse in our simulations}
\label{app:sectionrecollapse}
We found that for the most extreme encounter, the models are stripped and re-collapse to form a remnant.  The most extreme case we found was $10\,\mathrm{M}_\odot$ MAMS, $\beta=2.15$ model where the remnant formed after two weeks. 

\begin{figure}
    \centering
    \includegraphics[width=0.8\columnwidth]{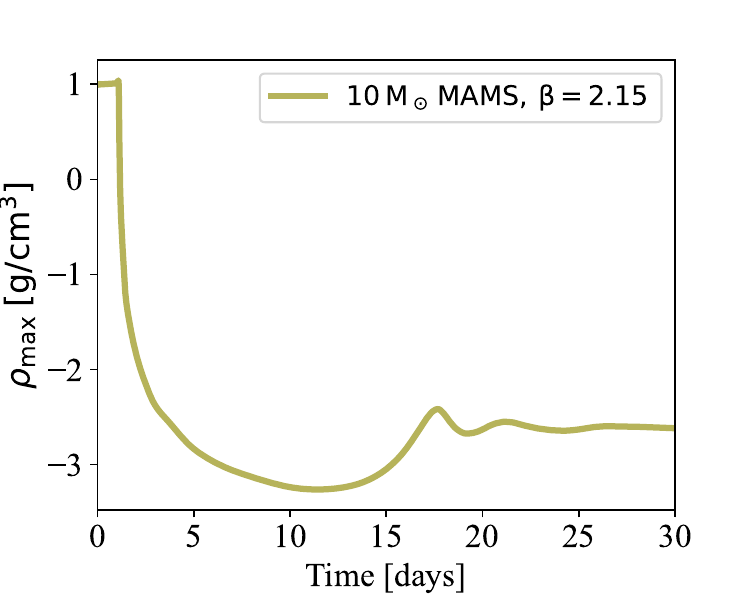}
    \caption{Maximum density as function of time for $10\,\mathrm{M}_\odot$ MAMS, $\beta=2.15$ model.  
 The central density decreases until $\sim15$ days and then increases, becoming constant at $\sim30$ days.}
    \label{fig:oscillation_beta_215}
\end{figure}

Figure~\ref{fig:oscillation_beta_215} shows the maximum density as a function of time for this model.  We note that the density decreases until $\sim15$ days, after which it increases, as the material re-collapses, leading to the formation of a remnant.  The central density becomes constant at about $30$ days, but the model does not look spherical. Hence, we used the snapshot at $60$ days for analysis. 

\section{Control case: \textsc{Kepler} mapping back}

\begin{figure*}
        \includegraphics[width=0.33\textwidth]{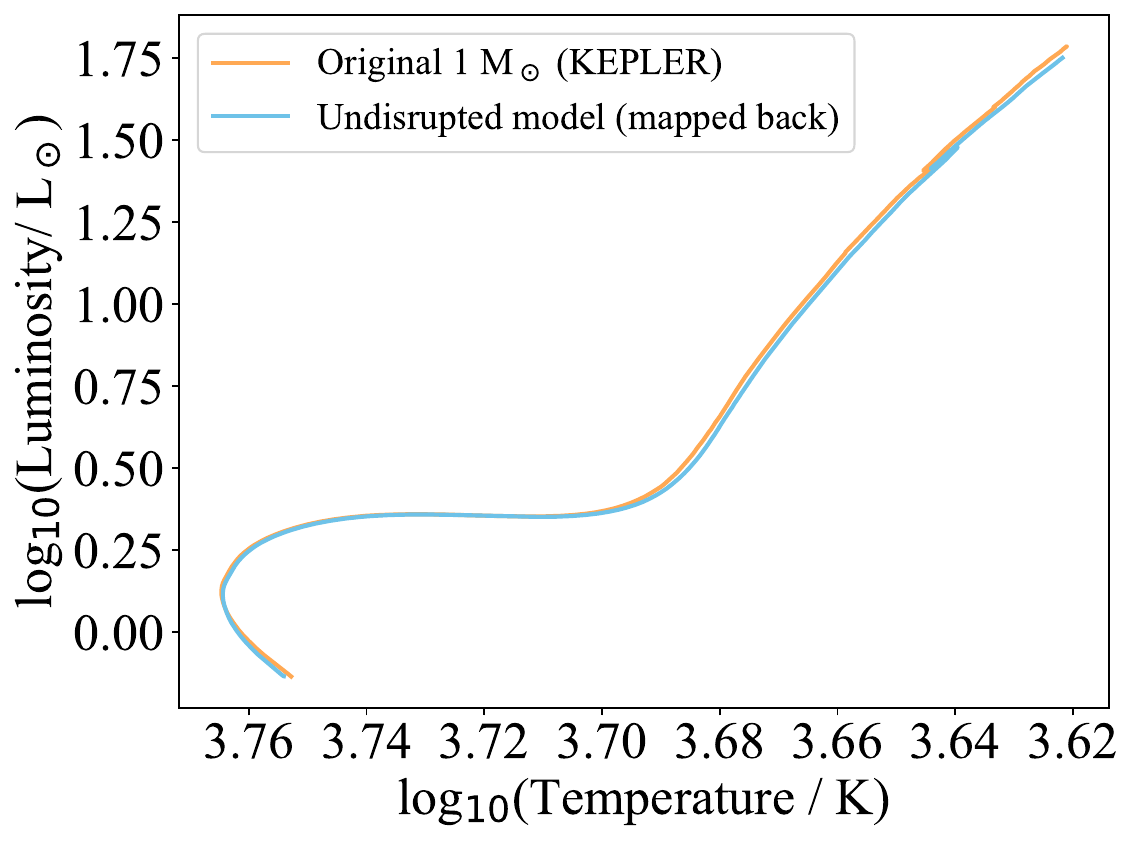}
    \includegraphics[width=0.33\textwidth]{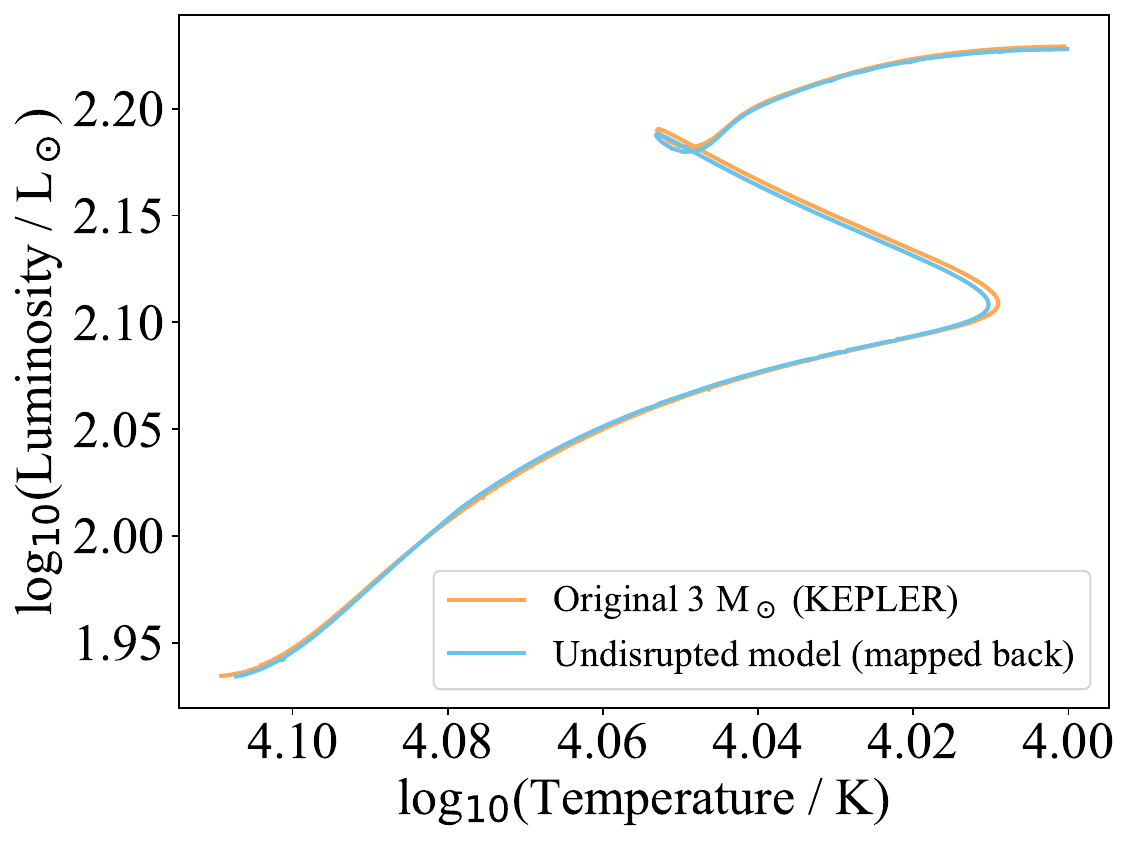}
     \includegraphics[width=0.33\textwidth]{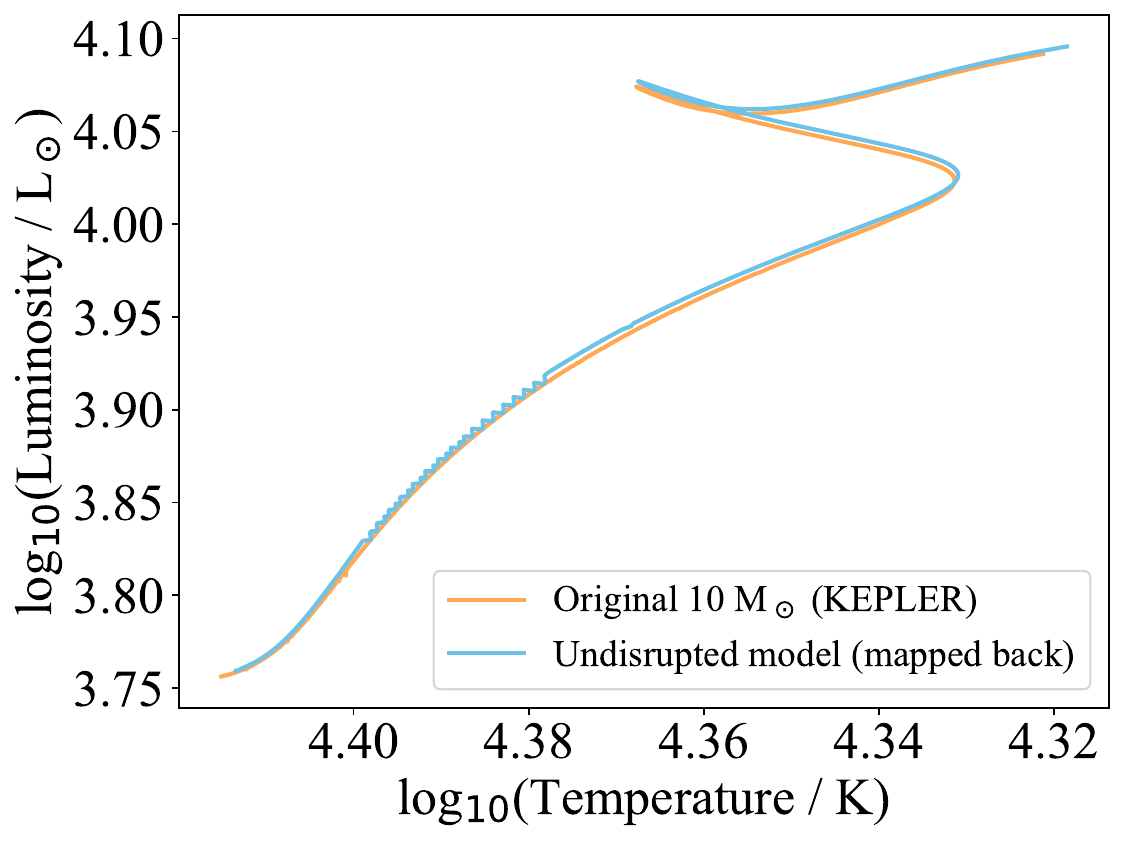}
    \caption{H-R diagrams of the original model compared with the mapped back model for  $1\,\mathrm{M}_\odot$, $3\,\mathrm{M}_\odot$, and $10\,\mathrm{M}_\odot$ (\textit{left} to \textit{right}) ZAMS models.  The \textsc{Kepler} model is shown in orange and the mapped back model is shown in blue.  We see that our mapped-back models are in good agreement with the original models. }
    \label{fig:hr_comparing_models}
\end{figure*}

To test that our \textsc{Kepler} mapping is robust, we compared the evolution of $1\,\mathrm{M}_\odot$, $3\,\mathrm{M}_\odot$, and $10\,\mathrm{M}_\odot$ ZAMS models that were mapped into \textsc{Phantom} but not disrupted. 
 Figure~\ref{fig:hr_comparing_models} shows the evolution of models in \textsc{Kepler} before they were mapped into \textsc{Phantom} (orange line), and models that were mapped back into \textsc{Kepler} from \textsc{Phantom} without disruption (blue line).  We can see that the $1\,\mathrm{M}_\odot$ mapped back model leaves the main-sequence at about at $99.8\%$ of the original model's time.  When the models reaches MAMS, the mapped back model has an luminosity of $0.97\,\mathrm{L}_\odot$ which is the same as the luminosity of the original model, but there is a difference in effective temperature.  The mapped back model has an effective temperature of $5\mathord,773\,\mathrm{K}$ while the original model has an effective temperature of $5\mathord,765\,\mathrm{K}$.  This corresponds to an error of $0.13\%$.  Hence, there are a few discrepancies. We found small differences in the mapping of the $3\,\mathrm{M}_\odot$ model. 

For $10\,\mathrm{M}_\odot$, mapped back model, we see oscillations but this effect is not significant enough for us to discard this mapping.  Overall, our mapped-back models are in good agreement with the original model's evolution, albeit small differences do exist.

\begin{figure}
    
    \includegraphics[width=\columnwidth]{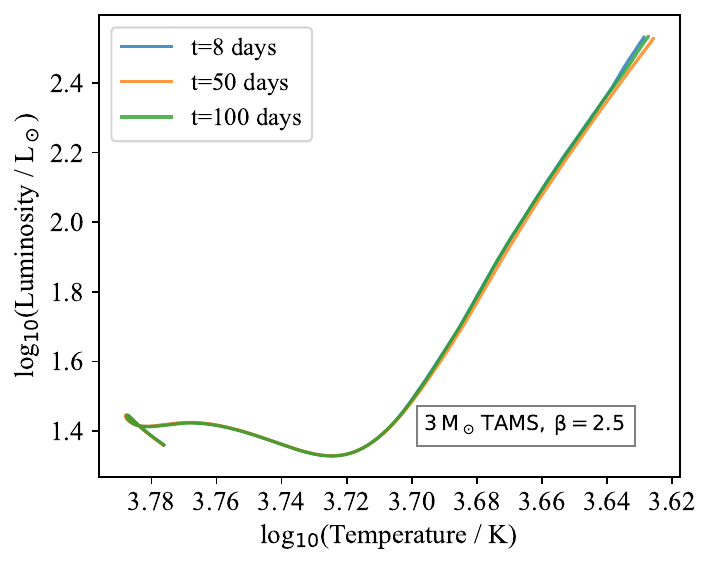}
    \caption{H-R diagram of $3\,\mathrm{M}_\odot$ TAMS model disrupted with $\beta=2.5$. We analysed snapshots at $8$ days, $50$ days, and $100$ days from the start of the simulation shown as blue, orange and green lines.  The results indicate good agreement in the star's evolution, suggesting that the time of analysis for the remnant snapshot does not significantly impact the outcome.}
    \label{fig:hr_time_effect}
\end{figure}
Figure~\ref{fig:hr_time_effect} shows the luminosity and effective temperature as a function of time of $3\,\mathrm{M}_\odot$ TAMS with $\beta=2.5$.  The model was mapped back at $8$ days, $50$ days, $100$ days.  We see that all models follow similar trends in luminosity and effective temperature as a function of time.  Hence, the time at which models are mapped back does not have a significant effect on stellar evolution.

\section{Implementing $E=0$ orbits in \textsc{Phantom}}
\label{app:zero_e_orbits}
\begin{table}
\centering
\caption{This table lists the simulation properties of hyperbolic and zero energy orbits for $1\,\mathrm{M}_\odot$ ZAMS models disrupted with different $\beta$ values.  The first column gives the orbit type that was set in \textsc{Phantom}. Column 2 lists the penetration factor. Column 3 lists the time of comparison between different orbits. Column 4 lists the mass of the remnant. Column 5 gives the estimated final velocity of the remnant at infinity based on its total energy. }
\begin{tabular}{cccccc}
\hline
 \hline 
Orbit type & $\beta$ & Time & $v_{\infty,\mathrm{ini}}$ & $v_{\infty,\mathrm{fin}}$ & $M_\mathrm{rem}$ \\
 & &  (days)  & ($\mathrm{km/s}$) & ($\mathrm{km/s}$) &  (M$_\odot$)\\
\hline

Hyperbolic & 0.54 & 1.68 & 2,034 & 1,948  & 1 \\
E=0 & 0.54 & 1.68 & 0 & (bound) & 1\\
\hline
Hyperbolic & 0.85 & 4 & 2,078 & 2,060 & 0.96 \\
E=0  & 0.85 & 4 & 0 & (bound) & 0.95\\
\hline
Hyperbolic & 1.06 & 3.25 & 2,093 & 2,067 & 0.83 \\
E=0  & 1.06 & 3.25 & 0 & (bound) & 0.81 \\
\hline
Hyperbolic & 1.49 & 8  & 2,109 & 2,205 & 0.15 \\
E=0  & 1.49 & 8  & 0 & 634 & 0.06\\
\hline
 \hline                                  
\end{tabular}

\label{tab:comparing_data}
\end{table}

\begin{table}
\centering
\caption{This table lists the semi-major axis and eccentricity of the bound models presented in Table~\ref{tab:comparing_data}.  Bound remnants were formed by disrupting the $1\,\mathrm{M}_\odot$ ZAMS model on $E=0$ orbits. Column 1 gives the impact parameter $\beta$. Column 2 lists the semi-major axis. Column 3 gives the eccentricity ($e$), specifically $\mathrm{log}_{10}(1-e)$. We assumed Keplerian orbits for our analysis.  }
\begin{tabular}{ccc}
\hline
 \hline 
$\beta$ & $a$ & $\mathrm{log}_{10}(1-e)$ \\

 & $(\mathrm{AU})$ &  \\
 \hline
 0.54 & $2.2\times10^3$ &-3.6  \\
 0.85 & $6.7\times10^3$ &-4.1 \\
 1.06 & $5.3\times10^3$ & -4.0 \\
\hline

\end{tabular}

\label{tab:ecc_a_two}
\end{table}

\begin{figure*}
    \centering
    \includegraphics[width=\textwidth]{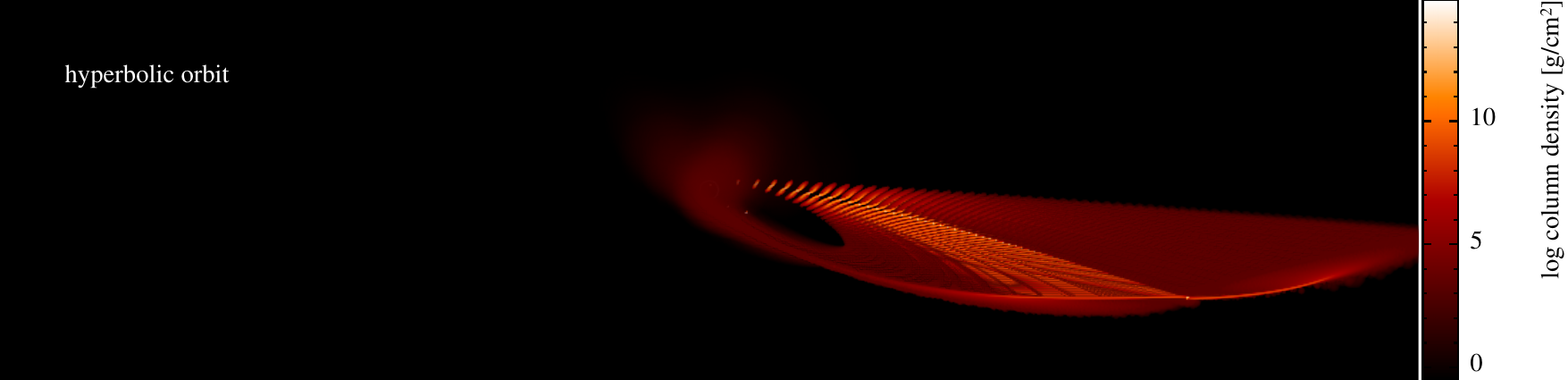}
    \includegraphics[width=\textwidth]{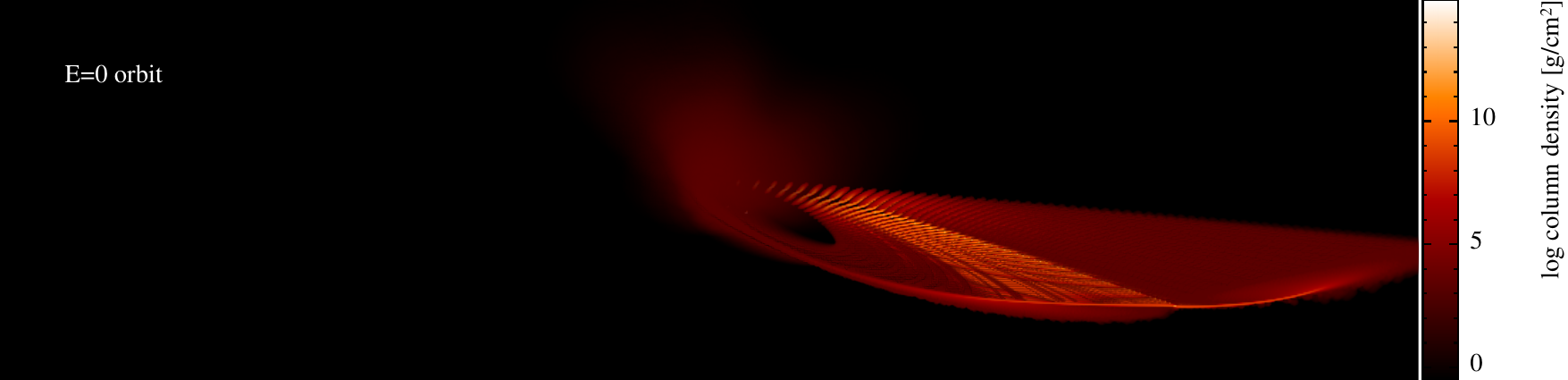}
  
         \caption{Snapshots of $1\, \mathrm{M}_\odot$ ZAMS disrupted, $\beta=1.49$. The \textit{top} panel shows the hyperbolic orbit and the \textit{bottom} panel corresponds to the zero energy orbit, showing the column density perpendicular to the orbital plane.  All simulations were run for $8$ days and each snapshot is at a $3$ hour interval. We note that orbits are similar, though zero energy orbit is slightly more perturbed, and has more mass being accreted onto the SMBH. }
    \label{fig:orbit_comparison}
\end{figure*}
% \begin{figure}
%     \centering
%     \includegraphics[width=\columnwidth]{omega_compa_zero.pdf}
%     \caption{Binned angular velocity as function of radius of $1$ $\mathrm{M}_\odot$ ZAMS model disrupted, $\beta=0.53$ and $\beta=1.49$ on hyperbolic and zero energy orbits. We see that the rotation profile of $\beta=0.53$ is similar but $\beta=1.49$ zero energy orbit has lower angular rotation than the hyperbolic orbit of same penetration factor.  }
%     \label{fig:rotation_ezero}
% \end{figure}
Here, we explain the appropriate way for implementing zero energy orbits in \textsc{Phantom}.  We used the geodesic code \citep{Liptai2019} to evolve $100$ test particles for a range of Newtonian pericentre distances ($47.6$--$196.8\,\mathrm{R}_\odot$).  but computed from a large distance of $2.4\times10^5\,\mathrm{R}_\odot$ where the Newtonian approximation is valid.  We then calculated the position and velocity of the particle at our starting distance of ten times the tidal radius for $1\,\mathrm{M}_\odot$ ZAMS model, once all the simulations were finished. This was achieved by interpolating to find the starting position and velocity closest to our desired initial separation in the geodesic trajectory.

%Subsequently, we utilised this function to determine the closest $t$ to ten tidal radius within a tolerance of $10^{-7}$.  The determined $t$ value was then employed to ascertain the velocity and position of the particle. We then performed B-spline interpolation between the $r_\mathrm{p}$ and corresponding position and velocity values at $10$ $r_\mathrm{t}$.  This helped us to interpolate position and velocity of the orbit for any $r_\mathrm{p}$ value. 

The resulting position and velocity were used for setting up orbit for the stars in \textsc{Phantom}. Figure~\ref{fig:orbit_comparison} shows the snapshots of $1\,\mathrm{M}_\odot$, $\beta=1.49$, showing column density perpendicular to the orbital plane. The \textit{top} panel shows a hyperbolic orbit and the \textit{bottom} panel shows a zero energy orbit. Both simulations were evolved $8$ days and each snapshot is at a $3$ day interval.  We see that the zero energy orbit is slightly more perturbed than the hyperbolic orbit. Moreover, the hyperbolic orbit remnant retains $2.5$ times higher mass than the zero energy orbit.

 Table~\ref{tab:comparing_data} lists the simulation results for both hyperbolic and zero energy orbits run for the same $\beta$ for comparison.  Column 1 gives the orbit type.  Column 2 gives the $\beta$ value.  Column 3 lists the time of comparison of models for the same $\beta$.  Columns 4 and 5 list the mass of the remnant and final escape velocity.  We see that the models have similar mass at lower $\beta$ values, but as the star gets closer to the SMBH, the remnant mass is lower for zero energy orbit compared with hyperbolic orbit for the same $\beta$.  Moreover, all remnants of zero energy orbits are bound except for the remnant of $\beta=1.49$ which has a velocity of $634\,\mathrm{km}\,\mathrm{s}^{-1}$.  This velocity might be the `kick' caused by the streams on the remnant as argued by \citet{Manukian2013}.

\label{lastpage}
\end{document}